\documentclass[12pt]{article}
\usepackage{amsmath}
\usepackage{amsfonts}
\usepackage{amssymb}
\usepackage{amsthm}
\usepackage{cite}

\newlength{\dinwidth}
\newlength{\dinmargin}
\newcommand{\Gadm}{\Ga_{M,\de}}
\newcommand{\xx}{\vx}
\newcommand{\te}{\textrm}
\newcommand{\alpbx}{\al^{\prime\prime\pm}}
\newcommand{\alppx}{\al^{\prime\pm}}
\newcommand{\allpx}{\al^{\pm}}
\newcommand{\almb}{\al^{\prime\prime-}}
\newcommand{\almp}{\al^{\prime-}}
\newcommand{\alm}{\al^{-}}
\newcommand{\alpb}{\al^{\prime\prime+}}
\newcommand{\alpp}{\al^{\prime+}}
\newcommand{\allp}{\al^{+}}
\newcommand{\whXi}{\widehat{\Xi}}
\newcommand{\alp}{\gamma}
\newcommand{\hF}{\widehat{F}}

\newcommand{\hXi}{\Xi}
\newcommand{\hPi}{\widehat{\Pi}}
\newcommand{\cA}{\mathring{A}}
\newcommand{\cB}{\mathring{B}}
\newcommand{\hB}{B}
\newcommand{\fin}{\mathcal{F}} %{\mathcal{F}}
\newcommand{\lin}{\mathcal{L}} %{\mathcal{L}}
\newcommand{\too}{\longrightarrow_{\su{\\ \!\!\!\!\!\!\!\!\!\!\!\!\!\! \K\to\infty}}}

\newcommand{\y}{\vec{y}}
\newcommand{\x}{\vx}
\newcommand{\nin}{\noindent}
\renewcommand{\Re}{\textrm{Re}}
\newcommand{\hmfa}{\mfa_0}
\newcommand{\si}{\sigma}
\newcommand{\M}{M_E}
\newcommand{\ube}{ \hat{\be}_{\su{\\ \!\!\!\!\!\! \mbox{\boldmath $_\gets$} }} }  %{\underline{\obe}}
\newcommand{\ual}{ \hat{\al}_{\su{\\ \!\!\!\!\!\! \mbox{\boldmath $_\to$} }} }%{\underline{\oal}}\hat{\al}_{\su{\\ \!\!\!\!\!\! 
\newcommand{\uall}{ \hat{\al}_{\su{\\ \!\!\!\!\!\! \mbox{\boldmath $_\gets$} }} }
\newcommand{\ualli}{ \hat{\al}_{\su{i \!\!\!\!\!\!\!\!\! \mbox{\boldmath $_\gets$} }} }
\newcommand{\uali}{ \hat{\al}_{\su{i \!\!\!\!\!\!\!\!\! \mbox{\boldmath $_\to$} }} }
\newcommand{\obe}{\hat{\beta}}
\newcommand{\oal}{\hat{\alpha}}

\newcommand{\umu}{\underline{\mu}}
\newcommand{\unu}{\underline{\nu}}
\newcommand{\uvx}{\underline{\vx}}
\newcommand{\Piz}{\Pi^{\prime}}%{\widetilde{\Pi}}
\newcommand{\Pizz}{\Pi^{\prime\prime}}

\newcommand{\rad}{r}
\newcommand{\Ppr}{P_{(p,\rad)} }
\newcommand{\ph}{\phantom}
\newcommand{\bx}{\underline{\x}}
\newcommand{\B}{(\mfa_{\cc}(\mco)^{\times N})^*} % {\mathcal{B}_N} % {(\mfa_{\cc}(\mco)^{\times N})^*}
\newcommand{\BB}{(\mfa_{\cc}(\mco)^{\times N-1})^* }
\newcommand{\BBB}{(\hmfa(\mco)^{\times M})^*}
\newcommand{\pp}{\textrm{pp}}
\newcommand{\cc}{\textrm{c}}

\newcommand{\bsharp}{\mbox{\boldmath $^\sharp$}}

\newcommand{\bflat}{\mbox{\boldmath $^\flat$}}

\newcommand{\Cs}{C_{\bsharp}}
\newcommand{\Csq}{C_{\bflat}} %{\Cs^{(q)}} \
\newcommand{\hQ}{\hat{Q}}
\newcommand{\he}{\hat{e}}

\newcommand{\R}{\textrm{Re}}
\newcommand{\I}{\textrm{Im}}

\newcommand{\Gad}{\Gamma_{N,\de}}
\newcommand{\Gadd}{\Gamma_{M,\de}}
\newcommand{\scc}{\textrm{c}}
\newcommand{\funi}{\fun_{\textrm{Im}} }
\newcommand{\funr}{\fun_{\textrm{Re}} }
\newcommand{\traceEB}{\trace_{E,1}}
\newcommand{\traceEBP}{\trace_{E,1}^+}

\newcommand{\K}{n}
\newcommand{\bnatural}{\mbox{\boldmath $^\natural$}}
\newcommand{\Nnat}{N_{\bnatural}}

\newcommand{\su}{\substack}
\newcommand{\N}{\mathcal{N}}

\newcommand{\fun}{\varphi}

\newcommand{\AN}{A_1\ot\cdots\ot A_N}

\newcommand{\ga}{\gamma}

\newcommand{\h}{\fr{1}{2}}

\newcommand{\fu}{h} %energy damping function.
\newcommand{\vx}{\vec{x}}

\newcommand{\nat}{\mathbb{N}}
\newcommand{\hil}{\mathcal{H}}
\newcommand{\om}{\omega}
\newcommand{\mfa}{\mathfrak{A}}
\newcommand{\mco}{\mathcal{O}}

\newcommand{\cone}{\overline{V}_+}

\newcommand{\trace}{\mathcal{T}}

\newcommand{\traceE}{\trace_E}

\newcommand{\La}{\Lambda}

\newcommand{\eps}{\varepsilon}

\newcommand{\fr}[2]{\frac{#1}{#2}}
\newcommand{\al}{\alpha}

\newcommand{\real}{\mathbb{R}}
\newcommand{\complex}{\mathbb{C}}

\newcommand{\ot}{\times}

\newcommand{\la}{\lambda}
\newcommand{\vp}{\phi}
\newcommand{\Fp}{f^+}
\newcommand{\Fm}{f^-}

\newcommand{\LL}{\mathcal{L}}
\newcommand{\ka}{\kappa}

\newcommand{\de}{\delta}
\newcommand{\non}{\nonumber}

\newcommand{\vxb}{\underline{\x}}

\newcommand{\vep}{\vec{p}}

\newcommand{\be}{\beta}

\newcommand{\Ga}{\Gamma}
\newcommand{\vac}{\Omega}

\newcommand{\Lp}{\mathcal{L}^{+}}
\newcommand{\Lm}{\mathcal{L}^{-}}
\newcommand{\Lpm}{\mathcal{L}^{\pm}}

\newcommand{\Epm}{\Lpm}
\newcommand{\LJ}{\mathcal{L}}

\newcommand{\mup}{\mu^+}
\newcommand{\mum}{\mu^-}

\newcommand{\half}{\fr{1}{2}}
\newcommand{\lan}{\langle}
\newcommand{\ran}{\rangle}

\def\proof{\noindent{\bf Proof. }}
\def\qed{$\Box$\medskip}
\newtheorem{theoreme}{Theorem } [section]
\newtheorem{proposition}[theoreme]{Proposition}
\newtheorem{lemma}[theoreme]{Lemma}
\newtheorem{definition}[theoreme]{Definition}
\newtheorem{corollary}[theoreme]{Corollary}
\newtheorem{remark}[theoreme]{Remark}
\newtheorem{example}[theoreme]{Example}
\newtheorem{criterion}[theoreme]{Criterion}

\newcommand{\beq}{\begin{equation}}
\newcommand{\eeq}{\end{equation}}
\newcommand{\beqa}{\begin{eqnarray}}
\newcommand{\eeqa}{\end{eqnarray}}
\newcommand{\ben}{\begin{arabicenumerate}}
\newcommand{\een}{\end{arabicenumerate}}
\newcommand{\bex}{\begin{example}}
\newcommand{\eex}{\end{example}}
\newcommand{\ber}{\begin{remark}}
\newcommand{\eer}{\end{remark}}
\newcommand{\bec}{\begin{corollary}}
\newcommand{\eec}{\end{corollary}}
\newcommand{\bep}{\begin{proposition}}
\newcommand{\eep}{\end{proposition}}
\newcommand{\becr}{\begin{criterion}}
\newcommand{\eecr}{\end{criterion}}

\def\bel{\begin{lemma}}
\def\eel{\end{lemma}}
\def\bet{\begin{theoreme}}
\def\eet{\end{theoreme}}
\def\bed{\begin{definition}}
\def\eed{\end{definition}}

\begin{document}
\title{Coincidence Arrangements of Local Observables and  Uniqueness of the Vacuum in QFT}
\author{ 
Wojciech Dybalski\\[5mm] 
  { Zentrum Mathematik,}
{ Technische Universit\"at M\"unchen,}\\ [2mm] 
{  D-85747 Garching, Germany}\\[2mm] 
e-mail: dybalski@ma.tum.de}
\date{}
\maketitle

\begin{abstract} 
A new phase space criterion, encoding the physically motivated behavior of 
coincidence arrangements of local observables, is proposed in this work.
This condition entails, in particular,  uniqueness and purity of the 
energetically accessible vacuum states. It is shown that the qualitative part
of this new criterion is equivalent to a compactness condition proposed in the 
literature. Its novel quantitative part is verified in massive free field theory.

\end{abstract}

\section{Introduction}
\setcounter{equation}{0}

Physical properties of  vacuum states have been a subject of study since the early days 
of algebraic quantum field theory \cite{BHS,Bor}. In particular, the problem of convergence of
physical states to a vacuum state under large translations attracted much attention. It was
considered under the assumptions of complete  (Wigner-) particle interpretation \cite{AH},
sharp mass hyperboloid \cite{BF}
and asymptotic abelianess in time \cite{Wanzenberg}. As none of these assumptions is expected to hold
in all  physically relevant models, further investigation of the vacuum structure is warranted.
We revisited this subject in  recent publications \cite{Dyb1,Dyb2}. 
There we proposed a phase space condition~$\Nnat$
which encodes the firm physical principle of additivity of energy over isolated subsystems. It entails the uniqueness of the vacuum states which can be prepared with a finite amount of energy. These vacuum states appear, 
in particular,  as  limits of physical states under large timelike translations in Lorentz covariant theories;
they can also be approximated by states of increasingly sharp energy-momentum values, in accordance with the uncertainty principle.

In the present paper  we introduce a new phase space condition~$\Csq$, stated below,  which 
is inspired by the fact that all elementary physical states are localized somewhere in space.
We show that this new criterion has all the physical consequences listed above and, in addition, entails purity of the vacuum state. A large part of the paper is devoted to the proof that the new  criterion holds in a model of massive, non-interacting particles and therefore is consistent with the basic postulates of quantum field theory \cite{Haag} which we now briefly recall.

The theory is based on a local net $\mco\to\mfa(\mco)$ of von Neumann algebras which are attached to open,
bounded regions of spacetime $\mco\subset\real^{s+1}$ and act on a Hilbert space $\hil$. The global algebra of 
this net, denoted by $\mfa$, is irreducibly represented on this space. Moreover, $\hil$ carries a strongly continuous unitary representation of the Poincar\'e group $\real^{s+1}\rtimes L_+^{\uparrow}\ni (x,\La)\to U(x,\La)$ which acts geometrically on the net
\beq
\al_{(x,\La)}\mfa(\mco)=U(x,\La)\mfa(\mco)U(x,\La)^{-1}=\mfa(\La\mco+x).
\eeq
We adopt the usual notation for translated operators $\al_x A=A(x)$ and functionals $\fun_{x}(A)=\fun(A(x))$,
where $A\in \mfa$,  $\fun\in \mfa^*$, and
demand that the joint spectrum of the generators $H, P_1,\ldots ,P_s$  of translations
is contained in the closed forward lightcone~$\cone$. We denote by $P_E$ the spectral projection of $H$ (the Hamiltonian) on the subspace spanned by vectors of energy lower than $E$.
Finally, we identify the predual of $B(\hil)$ with the space $\trace$ of trace-class operators on $\hil$
and  denote by $\traceE=P_E\trace P_E$ the set of normal functionals of energy bounded by $E$. The states
from $\mfa^*$ which belong to the weak$^*$ closure of $\traceEB$ for some $E\geq 0$ will be called the
energetically accessible states. (Here $\traceEB$ denotes the unit ball in the Banach space $\traceE$).

Important motivation for the present study comes from the refined spectral theory of translation automorphisms 
\cite{Dyb3}. The aim of this theory is to decompose the algebra of observables
$\mfa$ into subspaces which differ in their behavior under translation automorphisms $\real^{s+1}\ni x\to\al_x$. 
The first step is to identify the pure-point spectrum: Suppose that $A\in\mfa$ is an eigenvector of translation
automorphisms i.e.
\beq
\al_{x}A=e^{iqx}A, \quad x\in\real^{s+1}
\eeq
for some $q\in\real^{s+1}$. Then $A$ belongs to the center of $\mfa$, since by locality 
\beq
\|[A,B]\|=\lim_{|\vx|\to\infty}\|[\al_{\vx}A,B]\|=0,\quad\quad B\in\mfa.
\eeq
The irreducibility assumption ensures that the center of $\mfa$ consists only of
multiples of unity. Hence the pure-point subspace is given by
$\mfa_{\pp}=\{\, \be I\, |\, \be\in\complex\,\}$. Since we do not have a concept of
orthogonality in $\mfa$, it is not obvious how to choose the complementing continuous
subspace $\mfa_{\cc}$. Suppose, however, that there exists a distinguished projection $P_{\pp}$ from $\mfa$ 
onto $\mfa_{\pp}$. Then it is natural to set $\mfa_{\cc}=\ker\,P_{\pp}$ and there follows 
the decomposition
\beq
\mfa=\mfa_{\pp}\oplus\mfa_{\cc}.  \label{spectral-decomposition}
\eeq
Motivated by the Ergodic Theorem from the setting of groups of unitaries
acting on a Hilbert space, we attempt to construct such a projection
by averaging the automorphisms $\{\al_x\}_{x\in\real^{s+1}}$ over the group 
$\real^{s+1}$. Thus we consider the approximants
\beq
P_{\pp,L}(A)=\fr{1}{|K_L|}\int_{K_L} d^{s+1}x\,\al_{x}(A),\quad\quad A\in\mfa,
\eeq
where $K_L:=\{\,(x^0,\vx)\in\real^{s+1}\,|\, x^0\in [-L^{\eps},L^{\eps}], |\vx|\leq L\}$,
$0<\eps<1$ and the integrals, defined in the weak sense for any finite $L>0$, are elements of $B(\hil)$. 
By locality, the weak$^*$ limit points of the net $\{P_{\pp,L}(A)\}_{L>0}$ belong to the commutant of $\mfa$. 
Hence, by the irreducibility assumption, they are multiples of unity. With this
information at hand one easily obtains:
%%%%%%%%%%%%%%%%%%%%%%%%%%%%%%%%%%%%%%%%%%%%%%%%%%%%%%%%%%%%%%%%%%%%%%%%%%%%%%%%%%%%%%%%%%%%%%%%%%%
\bet Let $\{\om_L\}_{L>0}$ be a net of states on $\mfa$ given by the formula
\beq
\om_L(A):=\om(P_{\pp,L}(A)),\quad\quad\quad A\in\mfa, \label{state-averages}
\eeq
for some state $\om\in\trace$. Let $\om_0^{\ga}\in\mfa^*, \ga\in \mathbb{I}$ be the limit
points of this net in the weak$^*$ topology of $\mfa^*$ and $\{\, \om_{L_{\be}}\,|\, \be\in\mathbb{J}^{\ga}\,\}$ the corresponding approximating subnets. Then each such limit point $\om_0^{\ga}$ is a translationally invariant, energetically accessible state which is independent of the state $\om$.
\eet
%%%%%%%%%%%%%%%%%%%%%%%%%%%%%%%%%%%%%%%%%%%%%%%%%%%%%%%%%%%%%%%%%%%%%%%%%%%%%%%%%%%%%%%%%%%%%%%%%%%%%
%%%%%%%%%%%%%%%%%%%%%%%%%%%%%%%%%%%%%%%%%%%%%%%%%%%%%%%%%%%%%%%%%%%%%%%%%%%%%%%%
\nin It is a simple consequence of the above theorem that for any $\ga\in\mathbb{I}$ we obtain
a projection $P_{\pp}^{\ga}$ on the pure-point subspace $\mfa_{\pp}$ which has the following form
%%%%%%%%%%%%%%%%%%%%%%%%%%%%%%%%%%%%%%%%%%%%%%%%%%%%%%%%%%%%%%%%%%%%%%%%%%%%%%%
\beq
P_{\pp}^{\ga}(A):=w^*\te{-}\lim_{\be} P_{\pp,L_{\be}}(A)=\om_0^{\ga}(A)I,\quad\quad A\in\mfa.  %\fr{1}{|K_{L_{\be}}|}\int_{K_{L_{\be}}} d^{s+1}x\,\al_{x}(A).
\eeq
On physical grounds we expect that the translationally invariant, energetically accessible states $\om_0^{\ga}$
are vacuum states which all coincide. 
%In this case there also holds the uniqueness of the 
%decomposition~(\ref{spectral-decomposition}). I
In order to establish these facts, we amend the general postulates 
stated above by some physically motivated phase space conditions: First, we consider the maps $\Pi_E:\traceE\to\mfa(\mco)^*$ given by
\beq
\Pi_E(\fun)=\fun|_{\mfa(\mco)}.
\eeq
It was argued by Fredenhagen and Hertel in some unpublished work, quoted in \cite{BP},
that in physically meaningful theories these maps should satisfy the following condition:
%%%%%%%%%%%%%%%%%%%%%%%%%%%%%%%%%%%%%%%%%%%%%%%%%%%%%%%%%%%%%%%%%%%%%%%%%%%
\begin{enumerate}
\item[] \bf Condition \rm $\Cs$. The maps $\Pi_E$ are compact\footnote{
We adopt the restrictive definition of compactness from \cite{BP}: a map
is compact if it can be approximated in norm by finite-rank mappings. See Section 2
for details.}
for any  $E\geq 0$ and any double cone $\mco$.
\end{enumerate}
%%%%%%%%%%%%%%%%%%%%%%%%%%%%%%%%%%%%%%%%%%%%%%%%%%%%%%%%%%%%%%%%%%%%%%%%%%%
It is a consequence of this criterion that any energetically accessible
and translationally invariant state is a vacuum state, as physically expected. 
(See e.g. Theorem~2.2 (a) of \cite{Dyb1}). However, the uniqueness of these vacuum states
does not seem to follow from the above assumption. In order to settle this issue, we
introduce a strengthened variant of this criterion which is inspired by the behavior
of coincidence arrangements of local observables. For this purpose we pick one reference 
vacuum state $\om_0\in\{ \,\om_0^{\ga} \,|\, \ga\in\mathbb{I} \,\}$, define the
corresponding continuous subspace  $\mfa_{\cc}=\ker\,\om_0$ and the  local
continuous subspaces $\mfa_{\scc}(\mco)=\{\, A\in\mfa(\mco)\,|\, \om_0(A)=0\,\}$.
Next, for any double cone $\mco$ we introduce the Banach space 
$\B$ of $N$-linear forms on $\mfa_{\cc}(\mco)$, equipped  with the norm
\beq
\|\psi\|=\sup_{\su{ A_i\in\mfa_{\cc}(\mco)_{1} \\ i\in\{ 1,\ldots, N \}  } }|\psi(A_1\times\cdots\times A_N)|. 
\label{strangenorm}
\eeq
In order to control the minimal distance  between 
the regions in which the measurements are performed, we define the set of admissible translates of the region~$\mco$
\beq
\Gad=\{\,\bx=(\x_1,\ldots,\x_N)\in\real^{Ns} \, | \,  \forall_{t\in]-\de,\de[, i\neq j} \ \mco+\x_i\sim \mco+\x_j+t\hat{e}_0\,\},
\eeq
where the symbol $\sim$ indicates spacelike separation and $\hat{e}_0$ is the unit vector in the time direction.
For any $\bx\in\Gad$ and $\fun\in\traceE$ we introduce the following  elements of $\B$
\beq
\fun_{\bx}(\AN)=\fun(A_1(\x_1)\ldots A_N(\x_N)). \label{forms}
\eeq
Next, we consider the maps $\Pi_{E,N,\de}: \traceE\times\Gad\to\B$,  given by
\beq
\Pi_{E,N,\de}(\fun,\bx)=\fun_{\bx},
\eeq
which are linear in the first argument. Postponing the formal definitions of boundedness and compactness
for such maps to Section 2, we state a theorem which is at the basis of our investigation.
%%%%%%%%%%%%%%%%%%%%%%%%%%%%%%%%%%%%%%%%%%%%%%%%%%%
\bet\label{equivalence} A theory satisfies Condition $\Cs$ if and only if the
maps $\Pi_{E,N,\de}$ are compact for any $E\geq 0$, $N\in\nat$, $\de>0$ 
and double cone $\mco\subset\real^{s+1}$.
\eet
%%%%%%%%%%%%%%%%%%%%%%%%%%%%%%%%%%%%%%%%%%%%%%%%%%%%%%%%%%%%%%%%%%%%%%%%%%%%
\nin This result, whose proof is given in Section \ref{equi},  opens the possibility
to encode the physically expected behavior of coincidence arrangement of detectors
 into the phase space structure of a theory. We note that any functional from $\traceE$ should
describe systems with only a finite number of distinct localization centers. Indeed, in a 
theory of particles of mass $m>0$ the maximal number of such centers $N_0(E)$ is given, essentially, by
$\fr{E}{m}$. If the number of detectors $N$ is larger than $N_0(E)$, then at least one of them
should give no response and the result of the entire coincidence measurement should be zero. We formulate
this observation mathematically as a strengthened, quantitative variant of Condition $\Cs$:
%%%%%%%%%%%%%%%%%%%%%%%%%%%%%%%%%%%%%%%%%%%%%%%%%%%%%%%%%%%%%%
\begin{enumerate}
\item[ ] \bf Condition \rm $\Csq$
\item[(a)] The maps $\Pi_{E,N,\de}$ are compact for any $E\geq 0$, $N\in\nat$, $\de>0$ and double cone 
$\mco\subset\real^{s+1}$.
\item[(b)] For any $E\geq 0$ there exists  $N_0(E)\in\nat$ s.t. for any $N>N_0(E)$ the 
$\eps$-content\footnote{Let $V$, $W$ be Banach spaces and let $\Ga$ be some set. Then the $\eps$-content of a map 
$\Pi: V\times\Ga\to W$
is the maximal natural number $\N(\eps)$ for which there exist elements 
$(v_1,x_1),\ldots,(v_{\N(\eps)},x_{\N(\eps)})\in V_1\times\Ga$ s.t.
$\|\Pi(v_i,x_i)-\Pi(v_j,x_j)\|>\eps$ for $i\neq j$.} %in the norm topology of $\mfa(\mco)^*$ 
$\N(\eps)_{E,N,\de}$ of the map $\Pi_{E,N,\de}$ satisfies 
\beq
\lim_{\de\to\infty}\N(\eps)_{E,N,\de}=1
\eeq
for any $\eps>0$.
\end{enumerate}
%%%%%%%%%%%%%%%%%%%%%%%%%%%%%%%%%%%%%%%%%%%%%%%%%%%%%%%%%%%%%%%%
It is our main result that the reference  state 
$\om_0$, which enters into the definition of the maps $\Pi_{E,N,\de}$,
is the unique, energetically accessible vacuum state in theories complying with Condition~$\Csq$. 
Thus it defines a distinguished projection $P_{\pp}(\,\cdot\,)=\om_0(\,\cdot\,)I$ on $\mfa_{\pp}$
which fixes decomposition~(\ref{spectral-decomposition}).

Our paper is organized as follows: In Section 2  we show that the 
qualitative part (a) of Condition $\Csq$ is equivalent to Condition $\Cs$ 
(i.e. we prove Theorem~\ref{equivalence} stated above). In Section 3
we study physical consequences of  Condition $\Csq$ which
include  uniqueness and purity of the energetically accessible vacuum state 
as well as various preparation procedures for this state. In Section 4 we 
verify the quantitative part~(b) of the new criterion in massive free field theory. The more technical part of this discussion is postponed to the Appendices. The paper closes with brief Conclusions.

The results presented in this paper were included in the PhD thesis of the author \cite{Dyb3} completed at the
University of G\"ottingen.

%%%%%%%%%%%%%%%%%%%%%%%%%%%%%%%%%%%%%%%%%%%%%%%%%%%%%%%%%%%%%%%%%%%%%%
\section{Equivalence of Conditions $\Csq$(a) and $\Cs$}\label{equi}
%%%%%%%%%%%%%%%%%%%%%%%%%%%%%%%%%%%%%%%%%%%%%%%%%%%%%%%%%%%%%%%%%%%%%%
\setcounter{equation}{0}

In this section we show that Condition $\Csq$(a) is equivalent to
the existing Condition~$\Cs$ i.e. we prove Theorem \ref{equivalence} stated in the Introduction.

First,  we specify  the notions
of compactness which are used in the formulation of Conditions $\Cs$ and $\Csq$: 
Let $V$ and $W$ be Banach spaces and let $\lin(V,W)$ denote the space of
linear maps from $V$ to $W$ equipped with the standard norm. Let $\fin(V,W)$ denote
the subspace of finite-rank mappings. More precisely, any $F\in\fin(V,W)$  is
of the form $F=\sum_{i=1}^n\tau_i\, S_i$, where $\tau_i\in W$ and $S_i\in V^*$.
We say that a map $\Pi\in\lin(V,W)$ is compact, if it belongs to the closure 
of $\fin(V,W)$ in the norm topology of $\lin(V,W)$. This concept is used in Condition~$\Cs$.
To formulate a notion of compactness which is adequate for Condition $\Csq$, we need a more general framework:
Let $\Ga$ be a set and let $\lin(V\times\Ga,W)$ be the space of maps from $V\times\Ga$ to $W$, linear in the first argument, 
which are bounded in the norm
\beq
\|\Pi\|=\sup_{\su{v\in V_1\\ x\in\Ga} }\|\Pi(v,x)\|. %\quad \Pi\in\lin(V\times\Ga,W).
\eeq
The subspace of finite-rank maps $\fin(V\times\Ga,W)$ contains
all the maps of the form $F=\sum_{i=1}^n\tau_i\,S_i$, where $\tau_i\in W$, $S_i\in\lin(V\times\Ga,\complex)$.
We say that a map $\Pi\in\lin(V\times\Ga,W)$ is compact, if it belongs to the closure of $\fin(V\times\Ga,W)$
in the norm topology of $\lin(V\times\Ga,W)$.\\
%%%%%%%%%%%%%%%%%%%%%%%%%%%%%%%%%%%%%%%%%%%%%%%%%%%%%%%%%%%%%%%%%%%%%%%%%%%%%%%%%%%%%%
\bf Proof of  $\Csq$(a)$\Rightarrow\!\Cs$: \rm\\  Setting $N=1$ in Condition~$\Csq$(a),
we obtain, for any $\eps>0$, a finite-rank map $F\in\fin(\traceE\times\real^s,\mfa_{\cc}(\mco)^*)$
s.t.
\beq
\sup_{\su{(\fun,\vx)\in\traceEB\times\real^s \\  A\in\mfa_{\cc}(\mco)_1  }} |\Pi_{E,1,\de}(\fun,\vx)(A)-F(\fun,\vx)(A)|
\leq\eps.
\eeq
Noting that $\Pi_{E,1,\de}(\fun,\vx)=\Pi_E(\fun_{\vx})|_{\mfa_{\cc}(\mco)}$ and making use
of the fact that $\h(A-\om_0(A)I)\in\mfa_{\cc}(\mco)_1$ for any $A\in\mfa(\mco)_1$, we obtain
\beq
\sup_{\su{\fun\in\traceEB \\ A\in\mfa(\mco)_1}} 
|\Pi_{E}(\fun)(A)-\fun(I)\om_0(A)-F(\fun,0)(A-\om_0(A)I)|\leq 2\eps.
\eeq
Thus we can approximate the map $\Pi_{E}$ in norm with finite-rank mappings up to an arbitrary accuracy i.e.
this map is compact. \qed\\
%%%%%%%%%%%%%%%%%%%%%%%%%%%%%%%%%%%%%%%%%%%%%%%%%%%%%%%%%%%%%%%%%%%%%%%%%%%%%%%%%%%%%%%%%
The opposite implication is more interesting. It says that the restriction imposed by
Condition $\Cs$ on the number of states which can be distinguished by measurements with
singly-localized detectors limits also the number of states which can be discriminated 
by coincidence arrangements of such detectors. We start our analysis from the observation 
that  spatial distance between the detectors suppresses the energy transfer between them. 
The proof of the following lemma relies on methods from \cite{BY}. 
%%%%%%%%%%%%%%%%%%%%%%%%%%%%%%%%%%%%%%%%%%%%%%%%%%%%%%%%%%%%%%%%%%%%%%%%%%%%%%%%%%%%%%%%%%%%%%%%
\bel\label{mollifiers} Let $\de>0$, $\be>0$. Define the function $g:\,\, ]-\pi,\pi]\to \complex$
as follows
\beq
g(\vp)=\fr{\be}{\pi}\ln|\cot\fr{\vp+\alp}{2}\cot\fr{\vp-\alp}{2}|,
\eeq
where $\alp=2\arctan e^{-\fr{\pi\de}{2\be}}$. Then, for any pair of bounded operators
$A$, $B$ satisfying $[A(t),B]=0$ for $|t|<\de$ and any functional $\fun\in e^{-\be H}\trace e^{-\be H}$,
there holds the identity
\beq
\fun(AB)=\fun([A, \cB_\be]_+)+\fun(Ae^{-\be H}\hB_\be e^{\be H})+
\fun(e^{\be H}\hB_\be e^{-\be H} A),
\label{claim}
\eeq
where $[\,\cdot\, , \,\cdot\, ]_+$ denotes the anti-commutator and we made use of the fact that 
$\fun(e^{\be H}\,\,\cdot\,\,)$, $\fun(\,\,\cdot\,\, e^{\be H})$ are elements of $\trace$. 
Here $\cB_\be$ and  $\hB_\be$ are elements of $B(\hil)$ given by the (weak) integrals
\beqa
\cB_\be&=&\fr{1}{2\pi}\int_{0}^{\alp}d\vp \ B(g(\vp))+\fr{1}{2\pi}\int_{\pi-\alp}^{\pi}d\vp \ B(g(\vp)),
\label{Bb}\\
\hB_\be&=&\fr{1}{2\pi}\int_\alp^{\pi-\alp}d\vp \ B(g(\vp)),\label{Bb1}
\eeqa
where $B(g(\vp))=e^{ig(\vp)H}Be^{-ig(\vp)H}$.
\eel
%%%%%%%%%%%%%%%%%%%%%%%%%%%%%%%%%%%%%%%%%%%%%%%%%%%%%%%%%%%%%%%%%%%%%%%%%%%%%%%%%%%%%%%%%%%%%%%%%%%%%%%%%%%
\proof It suffices to prove the statement for functionals of the form $\fun(\,\cdot\,)=(\Psi_1|\,\cdot\,\Psi_2)$,
where $\Psi_1$ and $\Psi_2$ are vectors from the domain of $e^{\be H}$.
For $\de>0$ and $\be>0$ we define the set 
\beq
G_{\be,\de}=\{ \, z\in \complex \, | \, |\I z|<\be\, \} \backslash \{ \, z \, | \, \I z=0, |\R z|\geq\de\, \}
\eeq
and introduce the following function, analytic on $G_{\be,\de}$ and continuous at its
boundary
\beqa
h(z)=\left\{ \begin{array}{ll}
(\Psi_1|Ae^{izH}Be^{-izH}\Psi_2) &\textrm{ for } 0<\I z<\be \\
(\Psi_1|e^{izH}Be^{-izH}A\Psi_2) &\textrm{ for } -\be<\I z< 0 \\
(\Psi_1|AB(z)\Psi_2)=(\Psi_1|B(z)A\Psi_2) &\textrm{ for } \I z=0 \textrm{ and } |\R z|<\delta.
\end{array} \right.
\eeqa
We make use of the following conformal mapping from the unit
disc $\{\ w \ | \ |w|<1 \}$ to $G_{\be,\de}$ \cite{BY}
\beq
z(w)=\fr{\be}{\pi}\big\{\ln\fr{1+we^{i\alp}}{1-we^{i\alp}}-\ln\fr{1-we^{-i\alp}}{1+we^{-i\alp}}\big\}.
\eeq
Setting $w=re^{i\vp}$, $0<r<1$, we obtain from the Cauchy formula
\beq
h(0)=\fr{1}{2\pi}\int_{-\pi}^{\pi}d\vp \ h\big(z(re^{i\vp})\big). \label{Cauchy}
\eeq
Since $h(z)$ satisfies the following bound on the closure of $G_{\be,\de}$
\beq
|h(z)|\leq \|A\| \, \|B\| \, \|e^{\be H}\Psi_1\| \, \|e^{\be H}\Psi_2\|,
\eeq
we can, by the dominated convergence theorem, extend the path of
integration in relation~(\ref{Cauchy}) to the circle $r=1$. In this limit we have \cite{BY}
\beqa
\R\, z(e^{i\vp})&=& g(\vp),\\
\I\, z(e^{i\vp})&=&\left\{ \begin{array}{ll}
0 & \textrm{ if } |\vp|<\alp \textrm{ or } \pi-\vp<\alp \textrm{ or } \pi+\vp<\alp \\
\be & \textrm{ if } \alp<\vp <\pi-\alp \\
-\be & \textrm{ if } \alp<-\vp <\pi-\alp.
\end{array} \right.
\eeqa
Consequently, we obtain from (\ref{Cauchy})
\beqa
& &(\Psi_1|AB\Psi_2)\non\\
&=&\fr{1}{2\pi}\int_{0}^{\alp}d\vp\,(\Psi_1|[A,B(g(\vp))]_+\Psi_2)+\fr{1}{2\pi}\int_{\pi-\alp}^{\pi}d\vp\, (\Psi_1|[A,B(g(\vp))]_+\Psi_2)\non\\
&+&\fr{1}{2\pi}\int_\alp^{\pi-\alp}d\vp\,
\bigg(\!(\Psi_1|Ae^{-\be H}B(g(\vp))e^{\be H}\Psi_2)+(e^{\be H}\Psi_1|B(g(\vp))e^{-\be H}A\Psi_2)\!\bigg),\ \ \ \ \ \ \
\eeqa
what concludes the proof. \qed\\
%%%%%%%%%%%%%%%%%%%%%%%%%%%%%%%%%%%%%%%%%%%%%%%%%%%%%%%%%%%%%%%%%%%%%%%%%%%%%%%%%%%%%%%%%%
%%%%%%%%%%%%%%%%%%%%%%%%%%%%%%%%%%%%%%%%%%%%%%%%%%%%%%%%%%%%%%%%%%%%%%%%%%%%%%%%%%%%%%%%%%
In order to complete the proof of Theorem~\ref{equivalence}, we have to proceed
from the sharp energy bounds assumed in Condition~$\Cs$ to the exponential energy damping which is established
in Lemma~\ref{mollifiers}. Making use of the fact that $\traceE^*=P_EB(\hil)P_E$, Condition $\Cs$ can
be restated as a requirement that the maps $\whXi_{E}: \mfa(\mco)\to B(\hil)$, given by 
$\whXi_{E}(A)=P_EAP_E$,
are compact for any $E\geq 0$ and any double cone $\mco$. (See \cite{BP} for a similar discussion). 
With the help of the estimate
\beq
\|e^{-\be H}Ae^{-\be H}-e^{-\be H}P_EAP_Ee^{-\be H}\|\leq 2\|A\| e^{-\be E},
\eeq
one also concludes that the maps $\Xi_{\be}$ and $\Xi_{\be_1,\be_2}$ from $\lin(\mfa(\mco), B(\hil))$, defined as 
\beqa
\hXi_{\be}(A)&=&e^{-\be H}Ae^{-\be H}, \label{hXi}\\
\hXi_{\be_1,\be_2}(A)&=&e^{-\be_1 H}A_{\be_2}e^{-\be_1 H}, \label{hXi1}
\eeqa
are compact for any $\be,\be_1,\be_2>0$ and any double cone $\mco$. $A_{\be_2}$ is given by definition~(\ref{Bb1}).\\
%%%%%%%%%%%%%%%%%%%%%%%%%%%%%%%%%%%%%%%%%%%%%%%%%%%%%%%%%%%%%%%%%%%%%%%%%%%%
\bf Proof of $\Cs\Rightarrow\Csq$(a):\rm \\
For any $\be>0$ we introduce the auxiliary maps $\hPi_{\be,N,\de}\in\lin(\trace\times\Gad,\B)$  given by
\beq
\hPi_{\be,N,\de}(\fun,\vxb)(A_1\times\cdots\times A_N)=\fun(e^{-(N+\h)\be H}A_1(\vx_1)\ldots A_N(\vx_N)e^{-(N+\h)\be H}).
\eeq
They are related to the maps $\Pi_{E,N,\de}\in\lin(\traceE\times\Gad,\B)$ by the following identity,
valid for any $\fun\in\traceE$
\beq
\Pi_{E,N,\de}(\fun,\vxb)=\hPi_{\be,N,\de}(e^{(N+\h)\be H}\fun e^{(N+\h)\be H},\vxb). \label{PihPi}
\eeq 
In order to prove  compactness of the maps $\Pi_{E,N,\de}$, it suffices to verify that
the family of mappings $\{\hPi_{\be,N,\de}\}_{\be>0}$ 
%from $\lin(\traceE\times\Gad,\B)$
is \emph{asymptotically compact} in the following sense: There exists a family of finite-rank maps 
$\hF_{\be,N,\de}\in\fin(\trace\times\Gad,\B)$ s.t.
\beq
\lim_{\be\to 0}\|\hPi_{\be,N,\de}-\hF_{\be,N,\de}\|=0. \label{AC}
\eeq
If this property holds, then, by identity (\ref{PihPi}), the maps $\Pi_{E,N,\de}$ can be approximated
in norm as $\be\to 0$ by the finite-rank maps $F_{\be,N,\de}\in\lin(\traceE\times\Gad,\B)$ defined as
\beq
F_{\be,N,\de}(\fun,\vxb)=\hF_{\be,N,\de}(e^{(N+\h)\be H}\fun e^{(N+\h)\be H},\vxb).
\eeq
We establish  property (\ref{AC}) by induction in $N$:
For $N=1$ the statement follows from compactness of the map $\hXi_{\fr{3}{2}\be}$ given by (\ref{hXi}).
Next, we assume that the family  $\{\hPi_{\be,N-1,\de}\}_{\be>0}$ is asymptotically compact and prove that
$\{\hPi_{\be,N,\de}\}_{\be>0}$ also has this property. For this purpose we pick $\fun\in\trace_1$, 
$A_1,\ldots, A_N\in\mfa_{\cc}(\mco)_1$ and $\vxb\in\Gad$. Then $A_1(\x_1)\ldots A_{N-1}(\x_{N-1})$ and $A_N(\x_N)$ satisfy the assumptions of Lemma \ref{mollifiers} and  we obtain
\beqa
& &\hPi_{\be,N,\de}(\fun,\vxb)(A_1\times\cdots\times A_N)\non\\
& &=\fun(e^{-(N+\h)\be H}[A_1(\vx_1)\ldots A_{N-1}(\vx_{N-1}), \cA_{N,N\be}(\vx_N)]_+ e^{-(N+\h)\be H})\non\\
& &+\hPi_{\be,N-1,\de}\big(\{\hXi_{\h\be,N\be}(A_N)(\vx_N)\,\fun\, e^{-\be H}\},\vx_1,\ldots,\vx_{N-1}\big)
(A_1\times\cdots\times A_{N-1})\non\\
& &+\hPi_{\be,N-1,\de}\big(\{ e^{-\be H}\,\fun\,\hXi_{\h\be,N\be}(A_N)(\vx_N)\},\vx_1,\ldots,\vx_{N-1}\big)
(A_1\times\cdots\times A_{N-1}).\,\,\,\,\,\,\, \,\,\,\label{inductive-step}
\eeqa
The first term on the r.h.s. of (\ref{inductive-step}) satisfies 
\beq
|\fun(e^{-(N+\h)\be H}[A_1(\vx_1)\ldots A_{N-1}(\vx_{N-1}), \cA_{N,N\be}(\vx_N)]_+ e^{-(N+\h)\be H})|\leq \fr{2\alp(N\be)}{\pi},
\label{rest1}
\eeq
where we made use of  definition (\ref{Bb}). We recall from the statement
of Lemma~\ref{mollifiers} that $\alp(N\be)\to 0$ with $\be\to 0$. To treat the remaining terms, we make use of the induction hypothesis: It assures that there exist finite-rank mappings $\hF_{\be,N-1,\de}\in\fin(\trace\times\Ga_{N-1,\de},\BB)$ s.t.
\beq
\lim_{\be\to 0}\|\hPi_{\be,N-1,\de}-\hF_{\be,N-1,\de}\|=0. \label{decay1}
\eeq
Next, making use of compactness of the maps $\hXi_{\h\be,N\be}\in\lin(\mfa(\mco),B(\hil))$, given by formula~(\ref{hXi1}), we can find
a family of finite-rank mappings $F_{\be}\in\fin(\mfa(\mco),B(\hil))$ s.t.
\beq
\|\hXi_{\h\be,N\be}-F_{\be}\|\leq \fr{\be}{1+\|\hF_{\be,N-1,\de}\| }, \label{decay2}
\eeq
for any $\be>0$. Now the second term on the r.h.s. of (\ref{inductive-step}) can be rewritten as follows
\beqa
& &\hPi_{\be,N-1,\de}(\{\hXi_{\h\be,N\be}(A_N)(\vx_N)\,\fun\, e^{-\be H}\},\vx_1,\ldots,\vx_{N-1})\non\\
& &\ph{4444}=(\hPi_{\be,N-1,\de}-\hF_{\be,N-1,\de})\big(\{\hXi_{\h\be,N\be}(A_N)(\vx_N)\,\fun\, e^{-\be H}\},\vx_1,\ldots,\vx_{N-1}\big)\non\\
& &\ph{4444}+\hF_{\be,N-1,\de}\big(\{(\hXi_{\h\be,N\be}(A_N)-F_{\be}(A_N))(\vx_N)\,\fun\, e^{-\be H}\},\vx_1,\ldots,\vx_{N-1}\big)\non\\
& &\ph{4444}+\hF_{\be,N-1,\de}\big(\{F_{\be}(A_N)(\vx_N)\,\fun\, 
e^{-\be H}\},\vx_1,\ldots,\vx_{N-1}\big).\label{inductive-step1}
\eeqa
We obtain from relations (\ref{decay1}) and (\ref{decay2}) that the first two terms on the r.h.s. 
of equation~(\ref{inductive-step1}) tend to zero with $\be\to 0$ in the norm topology of $\BB$,
uniformly in  $\fun\in\trace_1$, $A_N\in\mfa_{\cc}(\mco)_1$ and $\vxb\in\Gad$.
The last term on the r.h.s. of relation~(\ref{inductive-step1}) coincides with the finite-rank  map 
$\hF_{\be,N,\de}^{(1)}\in\fin(\trace\times\Gad,\B)$, given by
\beqa
& &\hF_{\be,N,\de}^{(1)}(\fun,\vx)(A_1\times\cdots\times A_N)\non\\
& &\ph{444}=\hF_{\be,N-1,\de}(\{F_{\be}(A_N)(\vx_N)\,\fun\, e^{-\be H}\},\vx_1,\ldots,\vx_{N-1})(A_1\times\cdots\times A_{N-1}).\,\,\,\,\,\,\,\,\,\,\,
\eeqa
The last term on the r.h.s. of (\ref{inductive-step}) can be analogously approximated by the maps
$\hF_{\be,N,\de}^{(2)}\in\fin(\trace\times\Gad,\B)$ defined as
\beqa
& &\hF_{\be,N,\de}^{(2)}(\fun,\vx)(A_1\times\cdots\times A_N)\non\\
& &\ph{444}=\hF_{\be,N-1,\de}(\{e^{-\be H}\,\fun\,F_{\be}(A_N)(\vx_N)\} ,\vx_1,\ldots,\vx_{N-1})(A_1\times\cdots\times A_{N-1}).\,\,\,\,\,\,\,\,\,\,\,
\eeqa
Summing up, we obtain from (\ref{inductive-step}) and  (\ref{rest1}) that 
\beq
\lim_{\be\to 0}\|\hPi_{\be,N,\de}-\hF_{\be,N,\de}^{(1)}-\hF_{\be,N,\de}^{(2)}\|=0,
\eeq
what concludes the inductive argument and the proof of Theorem \ref{equivalence}. \qed
%%%%%%%%%%%%%%%%%%%%%%%%%%%%%%%%%%%%%%%%%%%%%%%%%%%%%%%%%%%%%%%%%%%%%%%%%%%%%%%%%%%%%%%%%%%%%%%%%%%%%%%%%%%%%%%%%%%%%
\section{Condition $\Csq$ and the Vacuum Structure}\label{vacuum-structure}
%%%%%%%%%%%%%%%%%%%%%%%%%%%%%%%%%%%%%%%%%%%%%%%%%%%%%%%%%%%%%%%%%%%%%%%%%%%%%%%%%%%%%%%%%%%%%%%%%%%%%%%%%%%%%%%%%%%%%%%%%%%%
\setcounter{equation}{0}

In this section we show that the vacuum state $\om_0$, which entered into
the formulation of Condition~$\Csq$, is pure and that it is the only energetically accessible vacuum
state. Similarly as in \cite{Dyb1}, there follows relaxation of physical states to $\om_0$ 
under large timelike translations and  appearance of this vacuum state as a limit of states 
of increasingly sharp energy-momentum values. 

We recall, that the $\eps$-contents of the maps $\Pi_{E,N,\de}$, which entered
into the formulation of Condition~$\Csq$, have  direct physical interpretation: They restrict
the number of different measurement results which may occur in  coincidence arrangements
of local operators from $\mfa_{\cc}(\mco)$. However, for many applications 
it is more convenient to work with the norms of the maps $\Pi_{E,N,\de}$. The link is
provided by the following lemma.
%%%%%%%%%%%%%%%%%%%%%%%%%%%%%%%%%%%%%%%%%%%%%%%%%%%%%%%%%%%%%%%%%%%%%%%%%%%%%%%%%%%%%%%%%%%%%%%
\bel\label{content} Let $V$ and $W$ be Banach spaces and let $\{\Ga_{\de}\}_{\de>0}$ be a family of sets
ordered by inclusion i.e. $\Ga_{\de_1}\subset\Ga_{\de_2}$ for $\de_1\geq\de_2$. Let
$\{\Pi_{\de}\}_{\de>0}$ be a  family of compact maps from $\lin(V\times\Ga_\de,W)$ and
let $\N(\eps)_{\de}$ be the respective $\eps$-contents.
Then there holds $\lim_{\de\to\infty}\N(\eps)_{\de}=1$ for any $\eps>0$ if and only if $\lim_{\de\to\infty}\|\Pi_{\de}\|=0$.
\eel
%%%%%%%%%%%%%%%%%%%%%%%%%%%%%%%%%%%%%%%%%%%%%%%%%%%%%%%%%%%%%%%%%%%%%%%%%%%%%%%%%%%%%%%%%%%%%%%%
\proof First, suppose that $\lim_{\de\to\infty}\N(\eps)_{\de}=1$ for any $\eps>0$. Since
the $\eps$-content takes only integer values, for any $\eps>0$ we can choose $\de_{\eps}$
s.t. $\N(\eps)_{\de}=1$ for $\de\geq\de_{\eps}$. Then, by definition of the $\eps$-content,
there holds for any $\de\geq\de_{\eps}$
\beq
\|\Pi_{\de}\|\leq\eps,
\eeq
what entails $\lim_{\de\to\infty}\|\Pi_{\de}\|=0$.

To prove the opposite implication, we proceed by contradiction: We recall that the $\eps$-content
of a compact map is finite for any $\eps>0$. Next, we note that for any fixed $\eps>0$ 
the function  $\de\to\N(\eps)_{\de}$ is decreasing and bounded from below by one, so there exists $\lim_{\de\to\infty}\N(\eps)_{\de}$. Suppose that this limit is strictly  larger than one. Then, 
by definition of the $\eps$-content, there exist nets $(\fun_1^{(\de)},\vx_1^{(\de)})$ 
and $(\fun_2^{(\de)},\vx_2^{(\de)})$ in $V_1\times\Ga_{\de}$ s.t.
\beq
\|\Pi_{\de}(\fun_1^{(\de)},\vx_1^{(\de)})-\Pi_{\de}(\fun_2^{(\de)},\vx_2^{(\de)})\|>\eps
\eeq
for any $\de>0$. However, this inequality contradicts the assumption that the norms
of the maps $\Pi_{\de}$ tend to zero with $\de\to\infty$. \qed\\
%%%%%%%%%%%%%%%%%%%%%%%%%%%%%%%%
With the help of the above lemma we reformulate Condition $\Csq$(b) as follows: For any $E\geq 0$ there exists
such natural number $N_0(E)$ that for any $N>N_0(E)$
\beq
\lim_{\de\to\infty}\sup_{\su{A_i\in\mfa_{\cc}(\mco)_{1} \\ i\in\{1,\ldots N\} \\ \vxb\in\Gad \ } }
\|P_E A_1(\x_1)\ldots A_N(\x_N) P_E\|= 0. \label{norms1}
\eeq 
We use this relation in the following key lemma.
%%%%%%%%%%%%%%%%%%%%%%%%%%%%%%%%%%%%%%%%%%%%%%%%%%%%%%%%%%%%%%%%%%%%%%%%%%%
\bel\label{key} Suppose that Condition $\Csq$ holds.  Then, for any $E\geq 0$, double cone $\mco$, 
and a sequence $\{\de(\K)\}_1^{\infty}$ s.t. $\de(\K)\too\infty$, %with $\K\to\infty$
the following assertions hold true:
\begin{enumerate}
\item[(a)] For any family of points $\{\x_i^{(\K)}\}_1^{\K}\in\Ga_{\K,\de(\K)}$ there holds 
\beq
\lim_{\K\to\infty}\sup_{\su{\fun\in\traceEB \\ A\in\mfa_{\cc}(\mco)_{1}}} \bigg|\fr{1}{\K}\sum_{i=1}^{\K}\fun(A(\x_i^{(\K)}))\bigg|= 0.\label{average}
\eeq
\item[(b)] For any unit vector $\he\in\real^s$, sequence $\{\la^{(\K)} \}_1^{\infty}\in\real_+$ and a family 
of  points $\{\x_i^{(\K)}\}_1^{\K}$,  s.t.
$\{\x_i^{(\K)}\}_1^{\K}\cup\{\x_i^{(\K)}+\la^{(\K)}\he\}_1^{\K}\in \Ga_{2\K,\de(\K)}$,
there holds
\beq
\lim_{\K\to\infty}\sup_{\su{\fun\in\traceEB \\ A,B\in\mfa_{\cc}(\mco)_{1}}} \bigg|\fr{1}{\K}\sum_{i=1}^{\K}\fun(A(\x_i^{(\K)}) B(\x_i^{(\K)}+\la^{(\K)}\he\ ))\bigg|=0.\label{average1}
\eeq
\end{enumerate}
\eel
%%%%%%%%%%%%%%%%%%%%%%%%%%%%%%%%%%%%%%%%%%%%%%%%%%%%%%%%%%%%%%%%%%%%%%%%%%%%%
\proof 
It is a well known fact that any normal, self-adjoint functional on a von Neumann algebra can be
expressed as a difference of two normal, positive functionals which are mutually orthogonal \cite{Sakai}.
It follows that any $\fun\in\traceEB$ can be decomposed as 
\beq
\fun=\funr^+-\funr^-+i(\funi^+-\funi^-), \label{decomp}
\eeq
where $\funr^\pm$, $\funi^\pm$ are positive functionals from $\traceEB$. Therefore, it suffices
to prove relations (\ref{average}) and (\ref{average1}) for the set $\traceEBP$ of positive functionals from $\traceEB$. 
By a similar argument, it suffices to consider self-adjoint operators $A,B\in\mfa_{\cc}(\mco)$
in both statements. 

We choose some positive functional $\fun\in\traceEBP$, pick  $m\in\nat$ s.t.
$N=2^m$ is sufficiently large to ensure that (\ref{norms1}) holds.
To prove (a), we  define the operators
$Q_{\K}=\fr{1}{\K}\sum_{i=1}^{\K}A(\x_i^{(\K)})$, $n\in\nat$, where $A\in\mfa_{\cc}(\mco)$ is self-adjoint,
assume that $\K\geq N$ and compute
\beqa
|\fun(Q_{\K})|^{N} &\leq& \fun(Q_{\K}^{N})=\fr{1}{\K^{N}}\sum_{i_1,\ldots, i_{N}}
\fun(A(\x_{i_1}^{(\K)})\ldots A(\x_{i_{N}}^{(\K)}))\phantom{4444444444}\non\\
&=&\fr{1}{\K^{N}}\sum_{\su{i_1,\ldots, i_{N} \\ \forall_{k\neq l} i_k\neq i_l  }}\fun(A(\x_{i_1}^{(\K)})\ldots A(\x_{i_{N}}^{(\K)}))
\non\\
&+& \fr{1}{\K^{N}}\sum_{\su{i_1,\ldots, i_{N} \\ \exists_{k\neq l} s.t. i_k=i_l}}
\fun(A(\x_{i_1}^{(\K)})\ldots A(\x_{i_{N}}^{(\K)}))\non\\
&\leq& 
\fr{1}{\K^{N}}\!\!\!\!\sum_{\su{i_1,\ldots, i_{N} \\ \forall_{k\neq l} i_k\neq i_l}}\!\!\!\|P_E A(\x_{i_1}^{(\K)})\ldots A(\x_{i_{N}}^{(\K)}) P_E\|+\fr{1}{\K^{N}}\!\!\!\!\sum_{\su{i_1,\ldots, i_{N} \\ \exists_{k\neq l} s.t. i_k=i_l}}\!\!\! \|A\|^{N}.
\eeqa
In the first step above we applied the Cauchy-Schwarz inequality and in the third step we extracted from the resulting
sum the terms in which all the operators are mutually spacelike separated. Clearly, there are 
${\K \choose N}N!\leq \K^N$ such terms. Therefore, the remainder (the last sum above) 
consists of
\beq
\K^N-{\K \choose N}N!\leq c_N \K^{N-1}
\eeq
terms. There follows the estimate
\beq
|\fun(Q_{\K})|^{N}\leq 
\sup_{\su{A\in\mfa_{\cc}(\mco)_{1} \\ (\x_1,\ldots,\x_N)\in\Ga_{N,\de(\K)} } } \|P_E A(\x_1)\ldots A(\x_N) P_E\|+ \fr{c_N}{\K}\|A\|^{N}, \label{norms}
\eeq
whose r.h.s. tends to zero with $\K\to\infty$ by (\ref{norms1}), uniformly in $\fun\in\traceEB$,  what concludes
the proof of (\ref{average}). 

\nin In order to prove (b) we proceed similarly: Let $\hQ_{\K}=\fr{1}{\K}\sum_{i=1}^{\K}A(\x_i^{(\K)})B(\x_i^{(\K)}+\la^{(\K)}\he)$,
where $A, B\in\mfa_{\cc}(\mco)$ are self-adjoint. Then,  for $\K\geq N$, we obtain
\beqa
& &|\fun(\hQ_{\K})|^{N}\leq \fun(\hQ_{\K}^{N})\non\\
&=&\fr{1}{\K^{N}}\sum_{i_1,\ldots, i_{N}}\fun(A(\x_{i_1}^{(\K)})B(\x_{i_1}^{(\K)}+\la^{(\K)}\he)\ldots A(\x_{i_{N}}^{(\K)})B(\x_{i_{N}}^{(\K)}+\la^{(\K)}\he)) \non\\
&=&\fr{1}{\K^{N}}\sum_{\su{i_1,\ldots, i_{N}\\ \forall_{k\neq l} i_k\neq i_l  }}
\fun(A(\x_{i_1}^{(\K)})B(\x_{i_1}^{(\K)}+\la^{(\K)}\he)\ldots A(\x_{i_{N}}^{(\K)})B(\x_{i_{N}}^{(\K)}+\la^{(\K)}\he))\non\\
&+&\fr{1}{\K^{N}}\sum_{\su{i_1,\ldots, i_{N}\\ \exists_{k\neq l} s.t. i_k=i_l}} \fun(A(\x_{i_1}^{(\K)})B(\x_{i_1}^{(\K)}+\la^{(\K)}\he)\ldots A(\x_{i_{N}}^{(\K)})B(\x_{i_{N}}^{(\K)}+\la^{(\K)}\he))\non\\
&\leq& 
\fr{1}{\K^{N}}\sum_{\su{i_1,\ldots, i_{N} \\ \forall_{k\neq l} i_k\neq i_l}}
\|P_E A(\x_{i_1}^{(\K)})B(\x_{i_1}^{(\K)}+\la^{(\K)}\he)\ldots A(\x_{i_{N}}^{(\K)})B(\x_{i_N}^{(\K)}+\la^{(\K)}\he)P_E\|\non\\
&+&\fr{1}{\K^{N}}\sum_{\su{i_1,\ldots, i_{N}\\ \exists_{k\neq l} s.t. i_k=i_l}} (\|A\|\, \|B\|)^{N}.
\eeqa
By the same reasoning as in case (a) we obtain the estimate
\beqa
|\fun(\hQ_{\K})|^{N} &\leq& \sup_{\su{A\in\mfa_{\cc}(\mco)_{1} \\ (\x_1,\ldots,\x_{2N})\in\Ga_{2N,\de(\K)} } } 
\|P_E A(\x_1)B(\x_2)\ldots A(\x_{2N-1})B(\x_{2N}) P_E\|\non\\
&+&\fr{c_N}{\K} (\|A\|\,\|B\|)^{N}.
\eeqa
By taking the limit $\K\to\infty$ we conclude the proof of (\ref{average1}). \qed\\
%%%%%%%%%%%%%%%%%%%%%%%%%%%%%%%%%%%%%%%%%%%%%%%%%%%%%%%%%%%%%%%%%%%%%%%%%%%%%%%%%%%%%%%%%%
Now we are ready to prove our main theorem.
%%%%%%%%%%%%%%%%%%%%%%%%%%%%%%%%%%%%%%%%%%%%%%%%%%%%%%%%%%%%%
\bet\label{main} Suppose that Condition $\Csq$ is satisfied. Then there hold the following assertions:
\begin{enumerate}
\item[(a)] Let $\om\in\mfa^*$ be a state in the weak$^*$ closure of $\traceEB$ which is invariant
under translations in space. Then $\om=\om_0$.
\item[(b)]  $\om_0$ is a pure state.
\end{enumerate}
\eet
%%%%%%%%%%%%%%%%%%%%%%%%%%%%%%%%%%%%%%%%%%%%%%%%%%%%%%%%%%%%%%%%%%%%%%%%%%%%%%%%%%%%%%%%%%%
\proof In part (a) we proceed similarly as in the proof of Theorem 2.2 (b) from \cite{Dyb1}:
Let $\{\fun_{\be}\}_{\be\in I}$ be a net of functionals from $\traceEB$ approximating $\om$ in the weak$^*$
topology  and let $A\in\mfa_{\cc}(\mco)$ i.e. $\om_0(A)=0$. We choose  families
of points $\{\x_i\}_1^{\K}$ in $\real^s$ s.t. $\{\x_i\}_1^{\K}\in\Ga_{\K,\de(\K)}$ for some 
sequence $\{\de(\K)\}_1^{\infty}$ which diverges to infinity with $\K\to\infty$. We note the
following relation
\beqa
|\om(A)|&=&|\fr{1}{\K}\sum_{i=1}^{\K}\om(A(\x_{i}^{(\K)}))|=\lim_{\be}|\fr{1}{\K}
\sum_{i=1}^{\K}\fun_{\be}(A(\x_{i}^{(\K)}))|\non\\
        &\leq&\sup_{\fun\in\traceEB}|\fr{1}{\K}\sum_{i=1}^{\K}\fun(A(\x_{i}^{(\K)}))|_{\su{ \\ \\ \ph{4} \K\to\infty }}
\!\!\!\!\!\!\!\!\!\!\!\!\!\!\longrightarrow 0,
\eeqa
where in the first step we made use of the fact that the state $\om$ is invariant under translations
in  space and in the last step we made use of Lemma \ref{key} (a). Since
local algebras are norm-dense in the global algebra $\mfa$, we conclude that $\ker\om_0\subset\ker\om$
and therefore the two states are equal.\\
Let us now proceed to part (b) of the theorem. In order to show purity of $\om_0$, it suffices to verify 
that for any $A, B\in\mfa_{\cc}(\mco)$, some
unit vector $\he\in\real^s$ and some sequence of real numbers $\{\la^{(\K)}\}_1^{\infty}$ s.t. 
$\la^{(\K)}\too\infty$ 
there holds
\beq
\lim_{\K\to\infty}\om_0(AB(\la^{(\K)}\he))=0. \label{clustering}
\eeq
To this end, we pick a net $\{\fun_{\be}\}_{\be\in I}$ of functionals from $\traceEB$, approximating $\om_0$ in the weak$^*$ topology. (Such nets exist, since $\om_0$ is energetically accessible). Next, we choose families of points $\{\x_i^{(\K)}\}_1^{\K}$ as in  part (b) of Lemma~\ref{key} and compute
\beqa
|\om_0(AB(\la^{(\K)}\he))|&=&|\fr{1}{\K}\sum_{i=1}^{\K}\om_0(A(\x_i^{(\K)})B(\x_i^{(\K)}+\la^{(\K)}\he))|\non\\
&=&\lim_{\be}|\fr{1}{\K}\sum_{i=1}^{\K}\fun_{\be}(A(\x_i^{(\K)})B(\x_i^{(\K)}+\la^{(\K)}\he))|\non\\
&\leq& \sup_{\fun\in\traceEB}|\fr{1}{\K}\sum_{i=1}^{\K}\fun(A(\x_i^{(\K)})B(\x_i^{(\K)}+\la^{(\K)}\he))|_{\su{ \\ \\ \ph{4} \K\to\infty }}
\!\!\!\!\!\!\!\!\!\!\!\!\!\!\longrightarrow 0,
\eeqa
what proves relation (\ref{clustering}). \qed\\
%%%%%%%%%%%%%%%%%%%%%%%%%%%%%%%%%%%%%%%%%%%%%%%%%%%%%%%%%%%%%%%%%%%%%%%%%%%%%%%%%%%%%%%%%%%%%%%%%%%%%%%%%%%%%%%%%%
As a corollary  we obtain the convergence of states of bounded energy to the vacuum state under large spacelike or timelike translations. (It is an interesting open problem if this corollary holds also for lightlike directions).
%%%%%%%%%%%%%%%%%%%%%%%%%%%%%%%%%%%%%%%%%%%%%%%%%%%%%%%%%%%%%%%%%%%%%%%%%%%%%%%%%%%%
\bec Let Condition $\Csq$ be satisfied. Then, for any state $\om\in\traceE$, $E\geq 0$, and 
a spacelike or timelike unit vector $\he\in\real^{s+1}$, there holds
\beq
\lim_{\la\to\infty}\om_{\la\he}(A)=\om_0(A) \textrm{ for } A\in\mfa.
\eeq
\eec
%%%%%%%%%%%%%%%%%%%%%%%%%%%%%%%%%%%%%%%%%%%%%%%%%%%%%%%%%%%%%%%%%%%%%%%%%%%%%%%%%%%%%%
\proof First, let $\he$ be a spacelike vector. Then, by locality, $\{A(\la\he)\}_{\la>0}$ is a
central net in $\mfa$ for any $A\in\mfa$. Thus, by the irreducibility assumption, its
limit points as $\la\to\infty$ in the weak$^*$ topology of $B(\hil)$ are multiples of the identity. It
follows that the limit points of the net $\{\om_{\la\he}\}_{\la>0}$ are translationally invariant and
energetically accessible states. By Theorem~\ref{main}~(a), the only such state is $\om_0$.

If $\he$ is a timelike vector,
the proof relies on an observation due to D. Buchholz that limit points of 
$\{\om_{\la\he}\}_{\la>0}$ as $\la\to\infty$ are invariant under translations in some 
spacelike hyperplane as a result of Lorentz covariance.
(See Lemma 2.3 of \cite{Dyb1}). Then it follows from Theorem 2.2 (a) of \cite{Dyb1} that these limit points are vacuum states. They coincide with $\om_0$ due to Theorem~\ref{main}~(a) above. \qed
%%%%%%%%%%%%%%%%%%%%%%%%%%%%%%%%%%%%%%%%%%%%%%%%%%%%%%%%%%%%%%%%%%%%%%%%%%%%%%%%%%%%%%%%%%%%%%

To conclude this survey of applications of Condition $\Csq$, we recall from \cite{Dyb1}
another physically meaningful procedure for the preparation of vacuum states: It is to construct states with 
increasingly sharp values of energy and momentum, and exploit the uncertainty principle.
Let $\Ppr$ be the spectral projection corresponding to the ball of radius $r$ centered around point $p$ in the
energy-momentum spectrum and $\trace_{(p,r)}=\Ppr\trace\Ppr$. Proceeding analogously as in Proposition~2.5 of \cite{Dyb1}
and exploiting relation~(\ref{average}), we obtain the following result: 
%%%%%%%%%%%%%%%%%%%%%%%%%%%%%%%%%%%%%%%%%%%%%%%%%%%%%%%%%%%%%%%%%%%%%%%%%%%%%
\bep\label{shrinking} Suppose that Condition $\Csq$ is satisfied. Then, for any $p\in\cone$ and
double cone $\mco$, there holds
\beq
\lim_{r\to 0}\sup_{\su{\fun\in\trace_{(p,r),1} \\ A\in\mfa(\mco)_{1}} }|\fun(A)-\fun(I)\om_0(A)|=0.
\eeq
\eep
%%%%%%%%%%%%%%%%%%%%%%%%%%%%%%%%%%%%%%%%%%%%%%%%%%%%%%%%%%%%%%%%%%%%%%%%%%%%%%
%%%%%%%%%%%%%%%%%%%%%%%%%%%%%%%%%%%%%%%%%%%%%%%%%%%%%%%%%%%%%%%%%%%%%%%%%%%%%%%%%%%%%%%%%%%
\section{Condition $\Csq$ (b) in Massive Scalar Free Field Theory} 
%%%%%%%%%%%%%%%%%%%%%%%%%%%%%%%%%%%%%%%%%%%%%%%%%%%%%%%%%%%%%%%%%%%%%%%%%%%%
\setcounter{equation}{0}

We showed in Section \ref{equi} that the qualitative part (a) of Condition $\Csq$ holds in
all theories satisfying Condition~$\Cs$, in particular in (massive and massless) scalar free field theory 
in physical spacetime \cite{BP}. Moreover, we argued in the Introduction that in physically meaningful,
massive theories there should also hold the strengthened, quantitative part (b) of this condition. As we
demonstrated in Section \ref{vacuum-structure}, this quantitative refinement
has a number of interesting consequences pertaining to the vacuum structure. It is the goal of the present
section to illustrate the mechanism which enforces Condition~$\Csq$ (b) by a direct computation in
the theory of massive, non-interacting particles.

To establish notation, we recall some basic facts concerning the free scalar field
theory in $s$ space dimensions: The Hilbert space $\hil$ is the Fock space over $L^2(\real^s, d^sp)$. To 
a double cone $\mco$, whose base is the $s$-dimensional ball $\mco_r$ of radius $r$, centered at the origin, 
there correspond the closed subspaces $\Lpm:=[\om^{\mp\fr{1}{2}}\widetilde{D}(\mco_r)]$ and
we denote the respective projections by the same symbols.
Defining $J$ to be the complex conjugation in  configuration space,
we introduce the real linear subspace
\beq
\LJ=(1+J)\Lp+(1-J)\Lm
\eeq
and the corresponding von Neumann algebra
\beqa
\mfa(\mco)=\{ \, W(f) \, | \, f\in\LJ \, \}^{\prime\prime},
\eeqa
where $W(f)=e^{i(a^*(f)+a(f))}$. The vacuum state $\om_0$ is induced by the Fock space
vacuum $\vac$, i.e. $\om_0(\,\cdot\,)=(\vac|\,\cdot\,|\vac)$, and there holds
\beq
\om_0(W(f))=e^{-\h\|f\|^2}.
\eeq 
The unitary representation of translations
has the following form in the single-particle space
\beq 
(U_1(x)f)(\vec{p})=e^{i(\om(\vep)t-\vep \vx)}f(\vec{p})=:f_x(\vec{p}), 
\eeq
where $x=(t,\vx)$, $\om(\vep)=\sqrt{\vep^2+m^2}$ and we assume that $m>0$. The translation
automorphisms are given by $\al_{x}(\,\cdot\,)=U(x)\,\cdot\,U(x)^*$, where $U(x)$
is the second quantization of $U_1(x)$. With the help of translations we define
local algebras attached to double cones centered at any point of spacetime.
Our task is to show that the resulting local net satisfies Condition~$\Csq$.
%%%%%%%%%%%%%%%%%%%%%%%%%%%%%%%%%%%%%%%%%%%%%%%%%%%%%%%%%%%%%%%%%%%%%
\bet\label{verification} Massive scalar free field theory satisfies Condition $\Csq$. 
\eet
%%%%%%%%%%%%%%%%%%%%%%%%%%%%%%%%%%%%%%%%%%%%%%%%%%%%%%%%%%%%%%%%%%%%%
\proof 
The main ingredient of the proof is the following elementary evaluation of the $N$-linear form
$\Pi_{E,N,\de}(\fun,\uvx)$, where  $\fun\in\traceEB$, $\uvx\in\Gad$, on the generating elements of $\mfa_{\cc}(\mco)$  
\beqa
& &\ph{444}\Pi_{E,N,\de}(\fun,\uvx)\big(\,\,\{W(f_1)-\om_0(W(f_1))I\}\times\cdots\times \{W(f_{N})-\om_0(W(f_N))I\}\, \big)\non\\
& &\ph{444444444}=\fun\big(\,(W(f_{1,\x_1})-\om_0(W(f_1))I)\ldots (W(f_{N,\x_N})-\om_0(W(f_N))I)\, \big)\non\\
& &\ph{444444444}=\sum_{R_1,R_2}(-1)^{|R_2|}e^{-\fr{1}{2}\sum_{k=1}^{|R_2|}\|f_{j_k}\|^2}
   \fun(W(f_{i_1,\x_{i_1}}+\cdots+f_{i_{|R_1|},\x_{i_{|R_1|}}}))\non\\
& &\ph{444444444}= \sum_{R_1,R_2} (-1)^{|R_2|}e^{-\fr{1}{2}\sum_{k=1}^N\|f_k\|^2} e^{-\sum_{1\leq k<l\leq|R_1|}
\Re\lan  f_{i_k,\x_{i_k}}|f_{i_l,\x_{i_l}}\ran} 
\non\\
& &\ph{444444444444444444444444444444444}\cdot\fun(:W(f_{i_1,\x_{i_1}})\ldots W(f_{i_{|R_1|},\x_{i_{|R_1|}}}):)\non\\
& &\ph{444444444} =\sum_{R_1,R_2}(-1)^{|R_2|}e^{-\fr{1}{2}\sum_{k=1}^N\|f_k\|^2}(e^{-\sum_{1\leq k<l \leq |R_1|}\Re\lan f_{i_k,\x_{i_k}}|f_{i_l,\x_{i_l}}\ran}-1)\non\\
& &\ph{444444444444444444444444444444444}\cdot\fun(:W(f_{i_1,\x_{i_1}})\ldots W(f_{i_{|R_1|},\x_{i_{|R_1|}}}):)\non\\
& &\ph{444444444}+e^{-\fr{1}{2}\sum_{k=1}^N\|f_k\|^2}\sum_{R_1,R_2}(-1)^{|R_2|}
\fun(:W(f_{i_1,\xx_{i_1}}+\cdots+f_{i_{|R_1|},\xx_{i_{|R_1|}} }):),\,\,\,\,\,\,\,\ \ \ \label{crucial}
%e^{-\fr{1}{2}\sum_{k=1}^N\|f_k\|^2} %e^{-\fr{1}{2} \|\uf\|^2}
%\fun\big(:(W(f_{1,\x_1})-I)\ldots (W(f_{N,\x_N})-I): \big), 
\eeqa
where the sum extends over all partitions $R_1=(i_1,\ldots,i_{|R_1|})$, $R_2=(j_1,\ldots,j_{|R_2|})$ of an
$N$-element set into two, possibly improper, ordered subsets. (If the condition $1\leq k<l\leq |R_1|$ is empty,
the corresponding sum is understood to be zero). In the second step we
made use of the fact that the Weyl operators are localized in spacelike
separated regions.
In the third step we applied the identity $W(f)=e^{-\h\|f\|^2}:W(f):$ and in the last step we 
divided the expression into two parts: The first part tends to zero for large spacelike separations, 
due to the decay of $\lan f_{1,x_1}|f_{2,x_2}\ran$ when $x_1-x_2$ tends to spacelike infinity.
In the next lemma we show that the last sum on the r.h.s. of (\ref{crucial}) vanishes for 
$N>2\fr{E}{m}$, so we can omit this last term in the subsequent discussion.
%%%%%%%%%%%%%%%%%%%%%%%%%%%%%%%%%%%%%%%%%%%%%%%%%%%%%%%%%%%%%%%%%%%%%%%%%%%%%%%%%%%%%%%%%%%%%%%%
\bel Let $E\geq 0$, $\fun\in\traceE$ and $N>2\fr{E}{m}$ be a natural number. Then there holds
\beq
S:=\sum_{R_1,R_2}(-1)^{|R_2|}\fun(:W(f_{i_1,\xx_{i_1}}+\cdots+f_{i_{|R_1|},\xx_{i_{|R_1|}} }):)=0,
\eeq
where the sum extends over all partitions of an $N$-element set into disjoint sets $R_1$, $R_2$.
\eel
%%%%%%%%%%%%%%%%%%%%%%%%%%%%%%%%%%%%%%%%%%%%%%%%%%%%%%%%%%%%%%%%%%%%%%%%%%%%%%%%%%%%%%%%%%%%%%%%%
\proof For any $f\in L^2(\real^s,d^sp)$ we introduce the map $M(f): B(\hil)\to B(\hil)$ given by
\beq
M(f)(C)=P_Ee^{ia^*(f)}Ce^{ia(f)}P_E,\quad\quad C\in B(\hil).
\eeq
The exponentials are defined by their Taylor expansions which are finite (in the massive theory) due
to the energy projections. The range of $M(f)$ belongs to $B(\hil)$ due to the energy
bounds \cite{BP} which in the massive case give
\beq
\|a(f_1)\ldots a(f_n)P_E\|\leq \bigg(\fr{E}{m}\bigg)^{\fr{n}{2}}\|f_1\|\ldots \|f_n\| \label{energy-bounds}
\eeq
for any $f_1,\ldots, f_n\in  L^2(\real^s,d^sp)$.
Making use of the fact that $e^{ia(f)}P_E=P_Ee^{ia(f)}P_E$, we obtain $M(f_1)M(f_2)=M(f_2)M(f_1)$  for any 
$f_1,f_2\in L^2(\real^s,d^sp)$. Moreover, $M(0)(C)=P_ECP_E$ for any $C\in B(\hil)$.
We denote by $\hat{I}$ the identity operator acting from $B(\hil)$ to $B(\hil)$. There
clearly holds
\beqa
S&=&\sum_{R_1,R_2}(-1)^{|R_2|}\fun\big(M(f_{i_1,\xx_{i_1}})\ldots M(f_{i_{|R_1|},\xx_{i_{|R_1|}} })(I)\big)\non\\
&=&\fun\big((M(f_{1,\xx_1})-\hat{I})\ldots (M(f_{N,\xx_N})-\hat{I})(I) \big)\non\\
&=&\fun\big((M(f_{1,\xx_1})-M(0))\ldots (M(f_{N,\xx_N})-M(0) )(I) \big), \label{positive-maps}
\eeqa
where the last equality holds due to the fact that $\fun\in\traceE$. Finally, we note that for any $C\in B(\hil)$
\beq
\big(M(f)-M(0)\big)(C)=\sum_{k+l\geq 1}P_E\fr{(ia^*(f))^k}{k!}C\fr{(ia(f))^l}{l!}P_E.
\eeq
Substituting this relation to (\ref{positive-maps}) the assertion follows. \qed\\
%%%%%%%%%%%%%%%%%%%%%%%%%%%%%%%%%%%%%%%%%%%%%%%%%%%%%%%%%%%%%%%%%%%%%%%%%%%%%%%%%%%%%%%%%%%%%%%%%
We will exploit relation~(\ref{crucial}) to show that for $N>2\fr{E}{m}$ the norms of the maps $\Pi_{E,N,\de}$
tend to zero with $\de\to\infty$, what entails Condition~$\Csq$ in view of Lemma~\ref{content}. To this end, we introduce
the $*$-algebra $\hmfa(\mco)$,  generated by finite linear combinations of Weyl operators, 
and denote by $\BBB$  the space of (not necessarily bounded) $M$-linear forms on $\hmfa(\mco)$. 
We define the maps $\Piz_{E,M,\de}:\traceE\times\Gadm\to\BBB$, linear in the first argument, extending
by linearity the following expression
\beqa
& &\Piz_{E,M,\de}(\fun,\uvx)(W(f_1)\times\cdots\times W(f_M))\non\\
& &=e^{-\fr{1}{2}\sum_{k=1}^M \|f_{k}\|^2} (e^{-\sum_{1\leq i<j \leq M}\Re\lan f_{i,\x_{i}}|f_{j,\x_{j}}\ran}-1)
\fun(:W(f_{1,\x_{1}}+\cdots+f_{M,\x_{M}} ):).\,\,\,\,\,\, \ \ \ \ \ \label{Piz}
\eeqa
We obtain from (\ref{crucial}) the equality valid for $N>2\fr{E}{m}$
\beqa
\Pi_{E,N,\de}(\fun,\uvx)\big(\,\{W(f_1)-\om_0(W(f_1))I\}\times\cdots\times \{W(f_{N})-\om_0(W(f_N))I\}\, \big)& &\non\\
=\sum_{R_1,R_2}
(-1)^{|R_2|}\om_0(W(f_{j_1}))\ldots\om_0(W(f_{j_{|R_2|}}))\cdot& &\non\\
\cdot\Piz_{E,|R_1|,\de}(\fun,\uvx)(W(f_{i_1})\times\cdots\times W(f_{i_{|R_1|} })).& & \label{crucial1}
\eeqa
To conclude the argument we need the following technical lemma.
%%%%%%%%%%%%%%%%%%%%%%%%%%%%%%%%%%%%%%%%%%%%%%%%%%%%%%%%%%%%%%%%%%%%%%%%%%%%
\bel\label{technical} For any $M\in\nat$, $E\geq 0$, double cone $\mco$ and sufficiently large $\de>0$ 
(depending on $M$, $E$ and $\mco$) there exist the maps 
$\Pizz_{E,M,\de}\in\lin(\traceE\times\Gadm,(\mfa(\mco)^{\otimes M})_*)$ which have the properties
\begin{enumerate}
\item[(a)] $\lim_{\de\to\infty}\|\Pizz_{E,M,\de}\|=0$,
\item[(b)] $\Pizz_{E,M,\de}(\fun,\vxb)(A_1\otimes\cdots\otimes A_M)=\Piz_{E,M,\de}(\fun,\vxb)(A_1\times\cdots\times A_M)$
for $A_1,\ldots, A_M\in\hmfa(\mco)$ and any $(\fun,\vxb)\in\traceE\times\Gadm$.
\end{enumerate}
\eel
%%%%%%%%%%%%%%%%%%%%%%%%%%%%%%%%%%%%%%%%%%%%%%%%%%%%%%%%%%%%%%%%%%%%%%%%%%%%%%
\noindent In view of this lemma, whose proof is postponed to
Appendix A, equality (\ref{crucial1}) can now be rewritten as follows, for sufficiently
large $N$, $\de$ and any $A_1,\ldots, A_N\in\mfa_{\cc}(\mco)$ 
\beqa
\Pi_{E,N,\de}(\fun,\uvx)(A_1\times\cdots\times A_N)=
\Pizz_{E,N,\de}(\fun,\uvx)(A_{1}\otimes\cdots\otimes A_{N}), \label{equality}
\eeqa
%%%%%%%%%%%%%%%%%%%%%%%%%%%%%%%%%%%%%%%%%%%%%%%%%%%%%%%%%%%%%%%%%%%%%%%%%%%%%%%%%%%%%%
where we made use of the facts that $\om_0(A_1)=\cdots=\om_0(A_N)=0$,  $\hmfa(\mco)$ is
dense in $\mfa(\mco)$ in the strong operator topology, and  $\Pizz_{E,M,\de}(\fun,\vxb)$ is a normal
functional on $(\mfa(\mco)^{\otimes M})$.
Consequently, for $N>2\fr{E}{m}$, the map $\Pi_{E,N,\de}$ shares the properties of 
$\Pizz_{E,N,\de}$ stated in  Lemma \ref{technical}. In addition we know from Theorem \ref{equivalence} that 
the maps $\Pi_{E,N,\de}$ are compact for any $\de>0$. Making use of Lemma \ref{content} 
we conclude that Condition $\Csq$ is satisfied. \qed\\
%%%%%%%%%%%%%%%%%%%%%%%%%%%%%%%%%%%%%%%%%%%%%%%%%%%%%%%%%%%%%%%%%%%%%%%%%%%%%%%%%%%%%%%
We remark that the assumption $m>0$ is used only in one (crucial) step in the above proof,
namely to eliminate the last term in relation (\ref{crucial}) and establish equality
(\ref{equality}). The properties of the maps $\Piz_{E,M,\de}$, stated in Lemma \ref{technical},
hold in massless free field theory as well. However, a complete verification argument for Condition~$\Csq$  
in the massless case has not been found yet.
%%%%%%%%%%%%%%%%%%%%%%%%%%%%%%%%%%%%%%%%%%%%%%%%%%%%%%%%%%%%%%%%%%%%%%%%%%%%%%%
\section{Conclusions}
%%%%%%%%%%%%%%%%%%%%%%%%%%%%%%%%%%%%%%%%%%%%%%%%%%%%%%%%%%%%%%%%%%%%%%%%%%%%%%%
\setcounter{equation}{0}

In this work we introduced and verified in a model the phase space condition~$\Csq$ which encodes 
localization of physical states in space. More precisely,
it says that any coincidence arrangement of spacelike separated
observables with vanishing vacuum expectation values gives zero response if the
number of observables is much larger than the number of localization centers which form the
state. From this physically motivated observation we derived detailed information
about the vacuum state: It is pure and unique in the energetically connected
component of the state space, it can be prepared with the help of states with
increasingly sharp energy-momentum values and appears as a limit of physical states under large
spacelike or timelike translations.

This last property corroborates the intuitive picture of  spreading of wave packets which
prevents the detection of particles with the help of observables of fixed spatial extension. In order to
determine the particle content and compute  collision cross sections, one has
to consider coincidence arrangements of particle detectors, whose responses are
suitably rescaled with time \cite{AH,BPS,Porr1,Porr2}. The methods developed in the present work
are also of relevance to the study of these interesting problems.

\bigskip

\noindent{\bf Acknowledgements:}
I would like to thank Prof. D. Buchholz for suggesting to me the present approach to the study of the
vacuum structure and for many valuable discussions in the course of this work.
Financial support from Deutsche Forschungsgemeinschaft and Graduiertenkolleg 
'Mathematische Strukturen in der modernen Quantenphysik' of the University of G\"ottingen 
is gratefully acknowledged.  This paper was completed at the TU M\"unchen under the DFG grant SP181/24.

\appendix
\section{Proof of Lemma \ref{technical}}
\setcounter{equation}{0}

The goal of this Appendix is to construct the maps $\Pizz_{E,M,\de}\in\lin(\traceE\times\Gadd,(\mfa(\mco)^{\otimes M})_*)$
and verify that they have the properties (a) and (b) specified in Lemma \ref{technical}. We will define these maps
as norm-convergent sums of rank-one mappings, i.e.
\beq
\Pizz_{E,M,\de}=\sum_{i=1}^{\infty}\tau_i\,S_i, \label{Pizz-expansion}
\eeq
where $\tau_i\in(\mfa(\mco)^{\otimes M})_*$ and $S_i\in\lin(\traceE\times\Gadd,\complex)$.

In order to construct a suitable family of functionals $\tau_i$, we recall some facts from
\cite{Bos3,Bos2}: Given any pair of multiindices $\mup,\mum$ and an orthonormal basis  $\{e_i\}_{1}^{\infty}$
of $J$-invariant eigenvectors in the single-particle space $L^2(\real^s,d^sp)$, one defines a normal functional $\tau_{\mup,\mum}\in B(\hil)_*$ by the following formula
\beqa
\tau_{\mup,\mum}(A)&\!=\!&
\bigg(\h\bigg)^{|\mup|+|\mum|}\!\!\!\!\!\!\!i^{-|\mup|-2|\mum|}\!\!\!\!\!\!
\sum_{\su{\allp+\alpp+\alpb=\mup\\ \alm+\almp+\almb=\mum}}\!\!\!\!\!\!\!\!
(-1)^{|\alpp|+|\almb|} 
\fr{\mup!}{\allp!\alpp!\alpb!}\fr{\mum!}{\alm!\almp!\almb!}\non\\
&\cdot&(\vac|a(e)^{\allp+\alm}a^*(e)^{\alpp+\almp}Aa^*(e)^{\alpb+\almb}\vac),
\eeqa
where $A\in B(\hil)$ and $\allpx, \alppx, \alpbx$ are multiindices. It is shown in
\cite{Bos3} that these functionals take the following values on the Weyl operators
\beq
\tau_{\mup,\mum}(W(f))=e^{-\half\|f\|^2}\lan e|\Fp\ran^{\mup}\lan e|\Fm\ran^{\mum}, \label{state}
\eeq 
where $f=f^++if^-\in\LL$ and $f^+$, $f^-$ are the real and imaginary parts of $f$ in configuration space.
It is also established there that the norms of these functionals satisfy the bound
\beq
\|\tau_{\mup,\mum}\|\leq 4^{|\mup|+|\mum|}(\mup!\mum!)^\half. \label{stateestimate}
\eeq
Turning to the definition of suitable functionals on $B(\hil)^{\otimes M}$, we introduce
$M$-tuples of  multiindices $\umu^{\pm}=(\mu^{\pm}_1,\ldots,\mu^{\pm}_M)$ and
the corresponding $2M$-multiindices $\umu=(\umu^+,\umu^-)$. We extend 
the standard rules of the multiindex notation as follows
\beqa
|\umu|&=&\sum_{i=1}^M(|\mu^+_i|+|\mu^-_i|),\\
\umu!&=&\prod_{i=1}^M\mu_i^+!\mu_i^-!,\\
\lan e | f \ran^{\umu}&=&\prod_{i=1}^{M}\lan e  | f_i^+  \ran^{\mu^+_i}\, \lan e | f_i^- \ran^{\mu^-_i},
\eeqa
where $f_i,\ldots, f_M\in \LL$. Now for any $2M$-multiindex $\umu$ we define a normal functional 
$\tau_{\umu}$ on $B(\hil)^{\otimes M}$ by the expression
\beq
\tau_{\umu}=\tau_{\mu_1^+,\mu_1^-}\otimes\cdots\otimes\tau_{\mu_M^+,\mu_M^-}. \label{Mtau}
\eeq
From relations (\ref{state}), (\ref{stateestimate}) and the polar decomposition of a normal functional
\cite{Sakai} one immediately obtains:
%%%%%%%%%%%%%%%%%%%%%%%%%%%%%%%%%%%%%%%%%%%%%%%%%%%%%%%%%%%%%%%%%
\bel\label{bound-tau} Let $\{e_i\}_1^{\infty}$ be an orthonormal basis in $L^2(\real^s,d^sp)$ of $J$-invariant
eigenvectors. The functionals $\tau_{\umu}\in (B(\hil)^{\otimes M})^*$ given by 
(\ref{Mtau}) have the following properties
\begin{enumerate}
\item[(a)] $\tau_{\umu}(W(f_1)\otimes\cdots\otimes W(f_M))=e^{-\half\sum_{k=1}^M\|f_k\|^2}\lan e|f\ran^{\umu}$, %\lan e|\Fm\ran^{\mum}$,
\item[(b)] $\|\tau_{\umu}\|\leq 4^{|\umu|}(\umu!)^\half$,
\end{enumerate}
where $f_1,\ldots,f_M\in \LL$.
\eel
%%%%%%%%%%%%%%%%%%%%%%%%%%%%%%%%%%%%%%%%%%%%%%%%%%%%%%%%%%%%%%%%%%%%%%%%%%%%%%%%%%%%%%%%%%%%%%%%%%%
In order to construct a basis $\{e_i\}_1^{\infty}$ in $L^2(\real^s,d^sp)$ of $J$-invariant eigenvectors, which is suitable for our purposes, we  recall, with certain  modifications, some material from the literature. Let $Q_E$ be the projection on states of energy lower than $E$ in the single-particle space. We define the operators $T_{E}^{\pm}=Q_E\Epm$
and $T_{\ka}^{\pm}=e^{-\fr{|\om|^{\ka}}{2}}\Epm$, where  $0<\ka<1$. 
By a slight modification of  Lemma~3.5  from \cite{BP} one finds that these operators satisfy
$\|T_E^{\pm}\|_1<\infty$, $\|T_{\ka}^{\pm}\|_1<\infty$, where
$\|\cdot\|_1$ denotes the trace norm. We define the operator $T$ as follows
\beq
T^2=|T_{E}^{+}|^2+|T_{E}^{-}|^2+|T_{\ka}^{+}|^2+|T_{\ka}^{-}|^2. \label{T}
\eeq
Making use of the estimate  $\|(A+B)^p\|_1\leq \|A^p\|_1+\|B^p\|_1$, valid for any $0<p\leq 1$ and any pair of positive operators $A$, $B$ s.t. $A^p$, $B^p$ are trace-class \cite{Kos}, we obtain
\beq
\|T\|_1\leq\|T_E^+\|_1+\|T_E^-\|_1 +\|T_{\ka}^+\|_1+\|T_{\ka}^-\|_1<\infty.  \label{lub3}
\eeq
Since $T$ commutes with $J$, it has a $J$-invariant orthonormal basis of eigenvectors $\{e_i\}_1^\infty$ 
and we denote the corresponding eigenvalues by $\{t_i\}_1^\infty$.

Now we proceed to the construction of the functionals $S_i\in\lin(\traceE\times\Gadd,\complex)$, to appear in  expansion
(\ref{Pizz-expansion}). Let $\oal^{\pm}=(\al_{1,2}^{\pm},\ldots,\al_{M-1,M}^{\pm})$ 
be  ${M \choose 2}$-tuples of multiindices and let $\oal=(\oal^+,\oal^-)$ be the corresponding $2{M \choose 2}$-multiindex.
First, we define the contribution to the functional which is responsible for the correlations between measurements:
\beqa
F_{\oal,\obe}(\uvx)
=\prod_{1\leq i<j\leq M}\fr{(-1)^{|\al_{i,j}^-|+|\al_{i,j}^+|}}{\sqrt{\al_{i,j}^+!\be_{i,j}^+!\al_{i,j}^-!\be_{i,j}^-!} }(\vac|a(\Lp e_{\vx_i})^{\al_{i,j}^+} a^*(\Lp e_{\vx_j})^{\be_{i,j}^+}\vac)\ph{4}& &\non\\
\ph{}\cdot(\vac|a(\Lm e_{\vx_i})^{\al_{i,j}^-} a^*(\Lm e_{\vx_j})^{\be_{i,j}^-}\vac),& & \label{F1}
\eeqa
where we use the short-hand notation $\Lpm e_{i,\vx_j}=U(\vx_j)\Lpm e_i$. The functionals in question are given by
\beqa
& &S_{\umu,\unu,\oal,\obe}(\fun,\uvx)=\fr{i^{|\umu^+|+|\unu^+|+2|\umu^-|} }{\umu!\unu!\sqrt{\oal!\obe!}}
F_{\oal,\obe}(\uvx)\fun(a^*(\LL e_{\vx})^{\umu}a(\LL e_{\vx})^{\unu}), \label{S1}
\eeqa
where $\fun\in\traceE$ and $\vxb\in\Gadd$.
The norms of these functionals satisfy the bound, stated in the following lemma, whose proof is postponed
to Appendix B.
%%%%%%%%%%%%%%%%%%%%%%%%%%%%%%%%%%%%%%%%%%%%%%%%%%%%%%%%%%%%%%%%%%%%%%%%%%%%%%%%%%%%%%%%%%%%%%%%%%%%%%%%%%%%%%%%%%%%%%
\bel\label{bound-S} The functionals $S_{\umu,\unu,\oal,\obe}\in\lin(\traceE\times\Gadd,\complex)$, given by (\ref{S1}),
satisfy the following estimates
\beq
\|S_{\umu,\unu,\oal,\obe}\|\leq\bigg(\fr{\M^{\fr{1}{2}(|\umu|+|\unu|)}}{\umu!\unu!}t^{\umu+\unu}\bigg)\bigg(\fr{1}{\sqrt{\oal!\obe!}}\sqrt{\fr{|\oal^+|!|\oal^-|!|\obe^+|!|\obe^-|!}{\oal!\obe!}} g(\de)^{|\oal|+|\obe|}t^{\oal+\obe}\bigg), \label{Sestimate1}
\eeq
where $\M=\fr{E}{m}$, $\{t_i\}_1^{\infty}$ are the eigenvalues of the operator $T$ given by (\ref{T}) and the function
$g$, which is independent of $\oal$ and $\obe$, satisfies $\lim_{\de\to\infty}g(\de)=0$.
\eel
%%%%%%%%%%%%%%%%%%%%%%%%%%%%%%%%%%%%%%%%%%%%%%%%%%%%%%%
Given the estimates from Lemmas \ref{bound-tau} (b) and \ref{bound-S} we can proceed to the study of
convergence properties of the expansion (\ref{Pizz-expansion}). For this purpose we need  some notation: For any
pair of ${M \choose 2}$-tuples of multiindices
$\oal^{\pm}=(\al_{1,2}^{\pm},\ldots,\al_{M-1,M}^{\pm})$ we define the associated $M$-tuples of multiindices $\ual^{\pm}$, $\uall^{\pm}$ as follows
\beqa
\uali^{\pm}&=&\sum_{\su{1<j\leq M \\ i<j}}\al_{i,j}^{\pm},\label{arrow1}\\
\ualli^{\pm}&=&\sum_{\su{1\leq j<M,\\ j<i}}\al_{j,i}^{\pm},\label{arrow2}
\eeqa
where $i\in\{1,\ldots, M\}$. The corresponding $2M$-multiindices are denoted by $\ual=(\ual^+,\ual^-)$, $\uall=(\uall^+,\uall^-)$.
The relevant estimate is stated in the following lemma, whose proof is given in Appendix B.
%%%%%%%%%%%%%%%%%%%%%%%%%%%%%%%%%%%%%%%%%%%%%%%%%%%%%%%%%%%%%%%%%%%%%%%%%%%%%%%%%%%%%%%%%%%%%%%%
\bel\label{convergence} The functionals $\tau_{\umu}\in(\mfa(\mco)^{\otimes N})_*$ and 
$S_{\umu,\unu,\oal,\obe}\in\lin(\traceE\times\Gadd,\complex)$ satisfy
\beq
\sum_{\su{\umu,\unu \\ \oal,\obe \\  (|\oal|,|\obe|)\neq (0,0) }} \|\tau_{\umu+\unu+\ual+\ube}\| \, \|S_{\umu,\unu,\oal,\obe}\|<\infty.
\eeq
for sufficiently large $\de>0$, depending on $M$, $E$ and the double cone $\mco$. Moreover, the above
sum tends to zero with $\de\to\infty$.
\eel
%%%%%%%%%%%%%%%%%%%%%%%%%%%%%%%%%%%%%%%%%%%%%%%%%%%%%%%%%%%%%%%%%%%%%%%%%%%%%%%%%%%%%%%%%%%%%%%%%%%
\nin After this preparation we  proceed to the main part of this Appendix.\\
%%%%%%%%%%%%%%%%%%%%%%%%%%%%%%%%%%%%%%%%%%%%%%%%%%%%%%
%%%%%%%%%%%%%%%%%%%%%%%%%%%%%%%%%%%%%%%%%%%%%%%%%%%%%
\nin\bf Proof of Lemma \ref{technical}: \rm\\
%%%%%%%%%%%%%%%%%%%%%%%%%%%%%%%%%%%%%%%%%%%%%%%%%%%%%%
We define  $\Pizz_{E,M,\de}\in\lin(\traceE\times\Gadd,(\mfa(\mco)^{\otimes M})_*)$ as follows 
\beq
\Pizz_{E,M,\de}(\fun,\vxb)=\sum_{\su{\umu,\unu \\ \oal,\obe \\  (|\oal|,|\obe|)\neq (0,0) }} \tau_{\umu+\unu+\ual+\ube}\, S_{\umu,\unu,\oal,\obe}(\fun,\vxb).
\eeq
In view of Lemma \ref{convergence} this map is well defined for sufficiently large $\de>0$ and satisfies
$\lim_{\de\to\infty}\|\Pizz_{E,M,\de}\|=0$ as required in part (a) of Lemma \ref{technical}. In order
to verify part (b), it suffices to show that for any $f_1,\ldots, f_M\in\LJ$
\beqa
\Pizz_{E,M,\de}(\fun,\vxb)(W(f_1)\otimes\cdots\otimes W(f_M))=\Piz_{E,M,\de}(\fun,\vxb)(W(f_1)\times\cdots\times W(f_M))& &\non\\
=e^{-\fr{1}{2}\sum_{k=1}^M \|f_{k}\|^2} 
(e^{-\sum_{1\leq i<j \leq M}
(\lan f_{i,\x_{i}}^+|f_{j,\x_{j}}^+\ran+\lan f_{i,\x_{i}}^-|f_{j,\x_{j}}^-\ran )}  -1) & &\non\\
\cdot\fun(:W(f_{1,\x_{1}}+\cdots+f_{M,\x_{M}} ):),& &\ \ \ \ \label{Piz1}
\eeqa
where the second equality restates the definition of the map $\Piz_{E,M,\de}$ given by formula (\ref{Piz}).
The l.h.s. can be evaluated making use of  Lemma \ref{bound-tau} (a) and definition~(\ref{S1}) 
\beqa
& &\Pizz_{E,N,\de}(\fun,\uvx)(W(f_1)\times\cdots\times W(f_N))\non\\
&=&\!\!\!\!e^{-\fr{1}{2}\sum_{k=1}^M \|f_{k}\|^2}\!\!\!\!\!\!\!\!\! \sum_{\su{ \umu,\unu \\ \oal, \obe \\ (|\oal|,|\obe|)\neq (0,0) }}\!\!\!\!\!\!\!\!\!\!\fr{i^{|\umu^+|+|\unu^+|+2|\umu^-|} }{\umu!\unu!\sqrt{\oal!\obe!}}
\lan e | f \ran^{\umu+\unu+\ual+\ube}F_{\oal,\obe}(\uvx)\fun(a^*(\LL e_{\vx})^{\umu}a(\LL e_{\vx})^{\unu}).
\,\,\,\,\,\,\,\,\, \ \ \ \ \label{Piz21}
\eeqa
First, we consider the sum w.r.t.  $\umu,\unu$. There holds 
\beqa
 \sum_{\su{ \umu,\unu }}\fr{i^{|\umu^+|+|\unu^+|+2|\umu^-|} }{\umu!\unu!}
\lan e | f \ran^{\umu+\unu}\fun(a^*(\LL e_{\vx})^{\umu}a(\LL e_{\vx})^{\unu})\ph{44444444444444444}\non\\
=\fun(:W(f_{1,\x_{1}}+\cdots+f_{M,\x_{M}} ):),
 \label{summunu}
\eeqa
as one can verify by expanding the normal ordered Weyl operator on the r.h.s. into the power series of
creation and annihilation operators of the functions $f_{j,\x_{j}}^{\pm}$, expanding each such function
in the orthonormal basis $\{e_i\}_1^{\infty}$ and making use of the multinomial formula 
\beq
a^{(*)}(f^{\pm}_{j,\x_j})^{m^{\pm}_j}=\sum_{\mu_j^{\pm},|\mu_j^{\pm}|=m^{\pm}_j}\fr{m_j^{\pm}!}{\mu_j^{\pm}!}\lan e | f_j \ran^{\mu_j^{\pm}}
a^{(*)}(\Lp e_{\x_j})^{\mu_j^{\pm}}.  \label{creation1}
\eeq
The sum w.r.t. $\oal,\obe$ in (\ref{Piz21}) gives
\beqa
\sum_{\su{\oal,\obe \\  (|\oal|,|\obe|)\neq (0,0) }} \fr{1}{\sqrt{\oal!\obe!}}\lan\, e | f \ran^{\ual+\ube}F_{\oal,\obe}(\uvx)
=\big(\!\prod_{1\leq i<j \leq M}e^{-\lan f_{i,\x_{i}}^+|f_{j,\x_{j}}^+\ran} e^{-\lan f_{i,\x_{i}}^-|f_{j,\x_{j}}^-\ran}\big) - 1.
\ \ \label{expo11} 
\eeqa 
This relation can be verified by expanding the exponential functions on the r.h.s. into Taylor series, making
use of the identity
\beq
\lan f_{i,\x_{i}}^{\pm}|f_{j,\x_{j}}^{\pm}\ran^{k^\pm_{i,j}}=
\fr{(\vac|a(f_{i,\x_{i}}^{\pm})^{k^\pm_{i,j}}a^*(f_{j,\x_{j}}^{\pm})^{k^{\pm}_{i,j}}\vac)}{k^{\pm}_{i,j}!}
\eeq
and applying to the resulting expressions expansions (\ref{creation1}). Comparing (\ref{expo11}) and (\ref{summunu}) with (\ref{Piz1}) we conclude the proof of Lemma \ref{technical}. \qed
%%%%%%%%%%%%%%%%%%%%%%%%%%%%%%%%%%%%%%%%%%%%%%%%%%%%%%%%%%%%%%%%%%%%%%%%%%%%%%%%%%%%%%%%%%%%%%%%%%%%%%%%%%%%%%%%
\section{Some Technical Proofs}
\setcounter{equation}{0}

In this Appendix we provide  proofs of Lemmas~\ref{bound-S} and \ref{convergence} which we
used in Appendix A to prove Lemma \ref{technical}.

The key ingredient of our proof of Lemma~\ref{bound-S} is
the observation that when the spatial distance between two local operators 
is large, then the energy transfer between them is heavily damped. We exploited
this idea in Section \ref{equi}, where it was encoded in Lemma \ref{mollifiers}.
In the present context it is more convenient to use the following result, which is a
 variant of Lemma~2.3 of \cite{universal}.
%%%%%%%%%%%%%%%%%%%%%%%%%%%%%%%%%%%%%%%%%%%%%%
\bel\label{damping} Let $\de>0$. Then there exists some continuous function
$\fu(\om)$, $\om\in\real$ which decreases almost exponentially, i.e.
\beq
\sup_{\om}|\fu(\om)|e^{|\om|^{\ka}}<\infty\textrm{ for } 0<\ka<1,
\eeq
%%%%%%%%%%%%%%%%%%%%%%%%%%%%%%%%%%%%%%%%%%%%%%
and which has the following property: For any pair of operators $A$, $B$ 
which have, together with their adjoints, a common, invariant, stable 
under time translations dense domain containing $\vac$ and satisfy $[A(t), B]=0$ for $|t|<\de$, there holds the identity
\beq
(\vac|AB\vac)=\half\big\{(\vac|A\fu(\de H)B\vac)+(\vac|B\fu(\de H)A\vac)\big\}.
\eeq
\eel
%%%%%%%%%%%%%%%%%%%%%%%%%%%%%%%%%%%%%%%%%%%%%%%%%%%%%%%%%%%%%%%%%%%%%%%%%%%%%%%
\nin With the help of this result we prove the following key lemma which will help us to control 
the correlation terms $F_{\oal,\obe}$.
%%%%%%%%%%%%%%%%%%%%%%%%%%%%%%%%%%%%%%%%%%%%%%%%%%%%%%%%%%%%%%%%%%%%%%%%%%%%%
\bel\label{damping1} Let $\de>0$, $(\x,\y)\in\Ga_{2,\de}$, $\{e_i\}_{1}^{\infty}$ be the basis of the $J$-invariant eigenvectors of the operator $T$ given by (\ref{T}), let $\{t_i\}_1^\infty$ be the corresponding eigenvalues  and let $\al,\be$ be multiindices. Then there holds, for any combination of $\pm$ signs,
\beq
|(\vac|a(\Lpm e_{\x})^{\al}a^*(\Lpm e_{\y})^{\be}\vac)|\leq \sqrt{|\al|!|\be|!} g(\de)^{|\al|+|\be|}
t^{\al+\be},\label{correlations} 
\eeq
where the function $g$  is independent of $\al$, $\be$ and satisfies $\lim_{\de\to\infty}g(\de)=0$.
\eel
%%%%%%%%%%%%%%%%%%%%%%%%%%%%%%%%%%%%%%%%%%%%%%%%%%%%%%%%
\proof We consider here only the (++) case, as the remaining cases are treated analogously.
We define the operators $\phi_{+}(e_i)=a^*(\Lp e_i)+a(\Lp e_i)$ and their
translates $\phi_{+}(e_i)(\x)=U(\x)\phi_{+}(e_i)U(\x)^{-1}$. Since the projection $\Lp$ commutes with
$J$ and $Je_i=e_i$, these operators are just the (canonical) fields of massive scalar free field theory. Since  $\de>0$,  locality
guarantees that $\phi_{+}(e_i)(\x)$ and $\phi_{+}(e_j)(\y)$ satisfy the assumptions of Lemma~\ref{damping}. Therefore,
we obtain
\beqa
& &\lan \Lp e_{i,\x}| \Lp e_{j,\y}\ran=(\vac|\phi_{+}(e_i)(\x)\phi_{+}(e_j)(\y)\vac)\non\\
& &\ph{444444}=\h\big((\vac|\phi_{+}(e_i)(\x)h(\de H)\phi_{+}(e_j)(\y)\vac)+(\vac|\phi_{+}(e_j)(\y)h(\de H)\phi_{+}(e_i)(\x)\vac)\big)\non\\
& &\ph{444444}=\fr{1}{2}\big(\lan \Lp e_{i,\x}|h(\de\om)\Lp e_{j,\y}\ran+\lan \Lp e_{j,\y}|h(\de\om)\Lp e_{i,\x}\ran\big).
\eeqa
Making use of this result, exploiting  the fact that the l.h.s. of (\ref{correlations}) vanishes for $|\al|\neq |\be|$
and setting $|\al|=|\be|=k$, we get
\beqa
(\vac|a(\Lp e_{\x})^{\al}a^*(\Lp e_{\y})^{\be}\vac)\ph{444444444444444444444444444444444444444}\non\\
=(\vac|a(\Lp e_{i_1,\x})\ldots a(\Lp e_{i_k,\x})a^*(\Lp e_{j_1,\y})\ldots a^*(\Lp e_{j_k,\y})\vac)& &\non\\
=\sum_{\si\in S_k}\lan \Lp e_{i_1,\x}| \Lp e_{j_{\si_1},\y}\ran \ldots \lan \Lp e_{i_k,\x}| \Lp e_{j_{\si_k},\y}\ran& &\non\\
=\sum_{\si\in S_k}\fr{1}{2}\big(\lan \Lp e_{i_1,\x}|h(\de\om)\Lp e_{j_{\si_1},\y}\ran+\lan \Lp e_{j_{\si_1},\y}|h(\de\om)\Lp e_{i_1,\x}\ran\big)& &\non\\   
\ldots \fr{1}{2}\big(\lan \Lp e_{i_k,\x}|h(\de\om)\Lp e_{j_{\si_k},\y}\ran+\lan \Lp e_{j_{\si_k},\y}|h(\de\om)\Lp e_{i_k,\x}\ran \big),
& &\quad\quad
\eeqa
where the sum extends over all permutations of a $k$-element set. For any $0<\ka<1$ there holds
$c_h^2:=\sup_{\om}|h(\om)e^{|\om|^{\ka}}|<\infty$. Consequently, we get
\beqa
|\lan \Lp e_{i,\x}|h(\de\om)\Lp e_{j,\y}\ran|&=&
|\lan \Lp e_{i,\x}| h(\de\om) e^{(\de|\om|)^{\ka}} e^{-(\de^{\ka}-1)|\om|^{\ka}}e^{-|\om|^{\ka}}\Lp e_{j,\y}\ran|\non\\
&\leq& c_{h}^2e^{-(\de^{\ka}-1) m^{\ka}} \|e^{-\fr{|\om|^\ka}{2}}\Lp e_i\|\,\|e^{-\fr{|\om|^\ka}{2}}\Lp e_j\|.
\eeqa
Finally, we note that 
$\|e^{-\fr{|\om|^\ka}{2}}\Lp e_i\|=\|T_{\ka}^+e_i\|\leq \|Te_i\|=t_i $
and the claim follows. \qed\\
%%%%%%%%%%%%%%%%%%%%%%%%%%%%%%%%%%%%%%%%%%%%%%%%%%%%%%%%%%%%%%%%%%%
After this preparation we proceed to the proof of Lemma~\ref{bound-S}.\\
%%%%%%%%%%%%%%%%%%%%%%%%%%%%%%%%%%%%%%%%%%%%%%%%%%%%%%%%%%%%%%%%%%%%%%%%%%%%%%%%%%%%%%%%%%%%%%%%%%%%%%%%%%
\bf Proof of Lemma~\ref{bound-S}:\rm\\
%%%%%%%%%%%%%%%%%%%%%%%%%%%%%%%%%%%%%%%%%%%%%%%%%%%%%%%%%%%%%%%%%%%%%%%%%%%%%%%%%%%%%%%%%%%%%%%%%%%%%%%%%%%%%%
Exploiting the energy bounds \cite{BP} (see estimate (\ref{energy-bounds}) above), we obtain
\beqa
|\fun(a^*(\LL e_{\vx})^{\umu}a(\LL e_{\vx})^{\unu})|&\leq& \M^{\fr{1}{2}(|\umu|+|\unu|)}\|Q_E\LL e\|^{\umu}\,
\|Q_E\LL e\|^{\unu}\non\\
&\leq& \M^{\fr{1}{2}(|\umu|+|\unu|)}t^{\umu+\unu}.\label{mubound1}
\eeqa
Next, with the help of  Lemma \ref{damping1} we analyze the expressions $F_{\oal,\obe}$ given by (\ref{F1})
\beqa
|F_{\oal,\obe}(\uvx)|\leq \prod_{1\leq i<j\leq M}\sqrt{\fr{|\al_{i,j}^+|!|\be_{i,j}^+|!|\al_{i,j}^-|!|\be_{i,j}^-|! }
{\al_{i,j}^+!\be_{i,j}^+!\al_{i,j}^-!\be_{i,j}^-!}}(g(\de)t)^{\al_{i,j}^++\be_{i,j}^++\al_{i,j}^-+\be_{i,j}^-}& &\non\\
\leq\sqrt{\fr{|\oal^+|!|\oal^-|!|\obe^+|!|\obe^-|!}{\oal!\obe!}} g(\de)^{|\oal|+|\obe|}t^{\oal+\obe},& &\label{Fbound1}
\eeqa
where we made use of the estimate $\prod_{1\leq i<j\leq M}|\al_{i,j}^+|!\leq (\sum_{1\leq i<j\leq M}|\al_{i,j}^+|)!=|\oal^+|!$. Altogether, combining (\ref{mubound1}) and (\ref{Fbound1}), we obtain from (\ref{S1}) the bound (\ref{Sestimate1}). \qed\\
%%%%%%%%%%%%%%%%%%%%%%%%%%%%%%%%%%%%%%%%%%%%%%%%%%%%%%%%%%%%%%%%%%%%%%%%%%%%%%%%%%%%%%%%%%%%%%%%%%%%%%%%%%%%%%%%
We conclude this Appendix with a proof of Lemma~\ref{convergence}.\\ 
%%%%%%%%%%%%%%%%%%%%%%%%%%%%%%%%%%%%%%%%%%%%%%%%%%%%%%%%%%%%%%%%%%%%%%%%%%%%%%%%%%%%%%%%%%%%%%%%%
\bf Proof of Lemma \ref{convergence}:\rm \\
%%%%%%%%%%%%%%%%%%%%%%%%%%%%%%%%%%%%%%%%%%%%%%%%%%%%%%%%%%%%%%%%%%%%%%%%%%%%%%%%%%%%%%%%%%%%%%%%%%
First, we estimate the norms of the functionals $\tau_{\umu+\unu+\ual+\ube}$.
Making use of the bound stated in Lemma~\ref{bound-tau}~(b)  and of the fact 
that $(a+b+c)!\leq 3^{a+b+c}a!b!c!$ for any $a,b,c\in\nat_0$, which follows from  properties of the multinomial 
coefficients, we get
\beqa
\|\tau_{\umu+\unu+\ual+\ube}\|&\leq& 4^{|\umu|+|\unu|+|\ual|+|\ube|}\sqrt{(\umu+\unu+\ual+\ube)!}\non\\
&\leq& \bigg( (4\sqrt{3})^{|\umu|+|\unu|}\sqrt{(\umu+\unu)!}\bigg)\bigg((4\sqrt{3})^{|\oal|+|\obe|} \sqrt{\ual!\ube!}\bigg), \label{prodstateestimate1}
\eeqa
where we noted that $|\ual|=|\oal|$ and $|\ube|=|\obe|$. (See definitions (\ref{arrow1}) and (\ref{arrow2})). 
The factor $\sqrt{ {\ual!\ube!} }$  in this bound will be controlled by the factor $\sqrt{\oal!\obe!}$
appearing in the denominator in (\ref{Sestimate1}). We note the relevant estimate
\beqa
\fr{\ual!}{\oal!}=\fr{\ual^+!}{\oal^+!}\fr{\ual^-!}{\oal^-!}
&=&\prod_{i=1}^M\fr{(\sum_{\su{1<j\leq M,\, j>i}}\al_{i,j}^+)!}
{(\prod_{1<j\leq M,\, j>i}\al_{i,j}^+)!}\fr{(\sum_{1<j\leq M,\, j>i}\al_{i,j}^-)!}{(\prod_{1<j\leq M,\, j>i}\al_{i,j}^-)!}
\non\\
&\leq& M^{\sum_{1\leq i<j\leq M}(|\al_{i,j}^+|+|\al_{i,j}^-|) }=M^{|\oal|}, \label{multiindexest1}
\eeqa
where we made use of properties of the multinomial coefficients. Similarly, the factor $\sqrt{(\umu+\unu)!}$
appearing in (\ref{prodstateestimate1}) will be counterbalanced by $\sqrt{\umu!\unu!}$ extracted from the denominator of (\ref{Sestimate1}). The relevant estimate relies on the property of the binomial coefficients
\beq
\fr{(\umu+\unu)!}{\umu!\unu!}\leq 2^{|\umu|+|\unu|}.
\eeq
With the help of the last two bounds and relations (\ref{Sestimate1}), (\ref{prodstateestimate1}) we obtain
\beqa
 \sum_{\su{\umu,\unu \\ \oal,\obe \\  (|\oal|,|\obe|)\neq (0,0) }} \|\tau_{\umu+\unu+\ual+\ube}\|\, \|S_{\umu,\unu,\oal,\obe}\|
\leq 
\sum_{\umu,\unu}\bigg(\fr{(4\sqrt{6\M})^{|\umu|+|\unu|} }{\sqrt{\umu!\unu!}}t^{\umu+\unu}\bigg)& & \non\\
\cdot\!\!\!\!\!\!\!\!\sum_{\su{\oal,\obe \\  (|\oal|,|\obe|)\neq (0,0) } }\!\!\!\!\!\!\!\!\bigg(\sqrt{\fr{|\oal^+|!|\oal^-|!|\obe^+|!|\obe^-|! }{\oal!\obe!}} (4\sqrt{3M}g(\de))^{|\oal|+|\obe|}t^{\oal+\obe}\bigg),& & \label{pnorm1}
\eeqa
where  we made use of the bound (\ref{multiindexest1}). The sum w.r.t. $\umu,\unu$ can
be easily estimated as it factorizes into $4M$ independent sums: Let $\mu$ be an ordinary multiindex, then
\beqa
\sum_{\umu,\unu}\bigg(\fr{(4\sqrt{6\M})^{|\umu|+|\unu|} }{\sqrt{\umu!\unu!}}t^{\umu+\unu}\bigg)
&=&\bigg(\sum_{\mu}\fr{(4\sqrt{6\M})^{|\mu|} }{\sqrt{\mu!}}t^{\mu}\bigg)^{4M}\non\\ 
&\leq&\bigg(\sum_{k=0}^{\infty}\fr{(4\sqrt{6\M})^{k} }{\sqrt{k!}}\sum_{\mu,|\mu|=k}\fr{|\mu|!}{\mu!}t^{\mu}\bigg)^{4M}\non\\
&\leq&\bigg(\sum_{k=0}^{\infty}\fr{(4\sqrt{6\M}\|T\|_1)^{k}}{\sqrt{k!}}\bigg)^{4M},
\eeqa
where in the second step we made use of the fact that the multinomial coefficients are greater than or equal to one
and in the last step we used  the multinomial formula. Clearly, the last sum is convergent. (As
a matter of fact it would suffice to consider $k\leq \M$ since $S_{\umu,\unu,\oal,\obe}$, given by formula
(\ref{S1}), vanishes for $|\umu|>\M$ or $|\unu|>\M$). As for the sum w.r.t. $\oal$, $\obe$ on the r.h.s. of
(\ref{pnorm1}), it suffices to study the case $|\oal^+|\neq 0$. Then the sum factorizes into four independent
sums  and we discuss here one of the factors
\beqa
& &\sum_{\oal^+,|\oal^+|\neq 0}\bigg(\sqrt{\fr{|\oal^+|!}{\oal^+!}}\bigg) (4\sqrt{3M}g(\de))^{|\oal^+|}t^{\oal^+}
\non\\
& &\ph{44444444444444}=\!\!\!\!\!\! \sum_{\su{\oal^+,|\oal^+|\neq 0 }}\!\!\!\!\!\!
(4\sqrt{3M}g(\de))^{|\oal^+|}\bigg(\sqrt{\fr{(|\al^+_{1,2}|+\cdots+|\al^+_{M-1,M}|)!}{\al^+_{1,2}!\ldots \al^+_{M-1,M}!}}\bigg)
t^{\oal^+}\non\\
& &\ph{44444444444444}\leq \sum_{\su{\oal^+, |\oal^+|\neq 0}}
(4\sqrt{3M^3}g(\de))^{|\oal^+|}\fr{|\al^+_{1,2}|!}{\al^+_{1,2}!}\ldots\fr{|\al^+_{M-1,M}|!}{\al^+_{M-1,M}!}t^{\oal^+}\non\\
& &\ph{44444444444444}\leq \sum_{\su{k^+_{1,2},\ldots,k^+_{M-1,M} \\ \sum_{1\leq i<j\leq M} k^+_{i,j}\neq 0}}
\prod_{1\leq i<j\leq M}(4\sqrt{3M^3}g(\de))^{k^+_{i,j}}\sum_{\al^+_{i,j},|\al^+_{i,j}|=k^+_{i,j}}\fr{|\al^+_{i,j}|!}{\al^+_{i,j}!}t^{\al^+_{i,j}}\non\\
& &\ph{44444444444444}\leq \sum_{\su{k^+_{1,2},\ldots,k^+_{M-1,M} \\ \sum_{1\leq i<j\leq M} k^+_{i,j}\neq 0}}(4\sqrt{3M^3}g(\de)\|T\|_1)^{(k^+_{1,2}+\cdots+k^+_{M-1,M})}. \label{last1}
\eeqa
In the second step we made use of the fact that 
\beq
\fr{(|\al^+_{1,2}|+\cdots+|\al^+_{M-1,M}|)!}{|\al^+_{1,2}|!\ldots |\al^+_{M-1,M}|!}\leq M^{2(|\al^+_{1,2}|+\cdots+|\al^+_{M-1,M}|)}
\eeq
and in the last step we exploited the multinomial formula. The last expression on the r.h.s. of (\ref{last1}) 
is a convergent geometric series for sufficiently large $\de$ and it tends to zero with $\de\to\infty$,  since $\lim_{\de\to\infty}g(\de)=0$. \qed\\

\end{document}